\documentclass[12pt]{article}

\usepackage{amssymb}
\usepackage{graphics}
\usepackage{makeidx}
\usepackage{layout}
\usepackage{amsmath}
\usepackage[vcentermath]{youngtab}
\let\a=\alpha

\newcommand{\beq}{\begin{equation}}
\newcommand{\eeq}{\end{equation}}
\newcommand{\beqn}{\begin{eqnarray}}
\newcommand{\eeqn}{\end{eqnarray}}

\newcommand{\nn}{\nonumber}

\newcommand{\ov}{\overline}

\newcommand{\be}{\begin{equation}}
\newcommand{\ee}{\end{equation}}
\newcommand{\ba}{\begin{eqnarray}}
\newcommand{\ea}{\end{eqnarray}}
\newcommand{\bdm}{\begin{displaymath}}
\newcommand{\edm}{\end{displaymath}}

\def\a{\alpha}

\newcommand{\ie}{{\it i.e.\ }}
\newcommand{\eg}{{\it e.g.\ }}

\DeclareMathAlphabet{\mathpzc}{OT1}{pzc}{m}{it}
\def\bea{\begin{eqnarray}}
\def\eea{\end{eqnarray}}
\def\beas{\begin{eqnarray*}}
\def\eeas{\end{eqnarray*}}
\def\sla{\raise.15ex\hbox{$/$}\kern-.57em}

\def\bea{\begin{eqnarray}}
\def\eea{\end{eqnarray}}
\def\de{\partial}

\def\sla{\raise.15ex\hbox{$/$}\kern-.57em}
\def\ie{{\it i.e.}~}
\def\eg{{\it e.g.}~}
\def\ap{{\alpha^\prime}}

\def\a{\alpha}

\def\cA{{\cal A}}
\def\cB{{\cal B}}

\def\cF{{\cal F}}

\def\cH{{\cal H}}
\def\cI{{\cal I}}
\def\cJ{{\cal J}}

\def\cN{{\cal N}}

\def\cU{{\cal U}}
\def\cV{{\cal V}}
\def\cW{{\cal W}}

\begin{document}
\begin{titlepage}
\begin{flushright}
{ROM2F/2010/17}
\end{flushright}
\vskip 1cm
\begin{center}
{\Large\bf On stable higher spin states
\\ \vspace{4mm} in Heterotic String Theories}\\
\end{center}
\vskip 2cm
\begin{center}
{\bf Massimo Bianchi}, {\bf Luca Lopez}, {\bf Robert Richter} \\
{\sl Dipartimento di Fisica, Universit\`a di Roma ``Tor Vergata''\\
I.N.F.N. Sezione di Roma ``Tor Vergata''\\
Via della Ricerca Scientifica, 00133 Roma, ITALY}\\
\end{center}
\vskip 1.0cm
\begin{center}
{\large \bf Abstract}
\end{center}

We study properties of 1/2 BPS Higher Spin states in heterotic
compactifications with extended supersymmetry. We also analyze non
BPS Higher Spin states and give explicit expressions for physical
vertex operators of the first two massive levels. We then study
on-shell tri-linear couplings of these Higher Spin states and
confirm that BPS states with arbitrary spin cannot decay into
lower spin states in perturbation theory. Finally, we consider
scattering of vector bosons off higher spin BPS states and extract
form factors and polarization effects in various limits.



\vfill

\end{titlepage}

\section*{Introduction}

One of the remarkable properties of String Theory -- probably its
hallmark -- is the presence of an infinite tower of massive higher
spin (HS) excitations in the free spectrum. Most of these are
unstable and can decay into lower spin states after turning on
interactions \cite{Chialva:2004xm, Iengo:2006gm, Chialva:2009pf,
Chialva:2009pg}. Some are long-lived and can be detected at
accessible energies if the string scale is much lower than the
Planck scale \cite{Dudas:1999gz, Chialva:2005gt, Bianchi:2006nf,
Lust:2008qc, Anchordoqui:2009mm, Dong:2010jt, Feng:2010yx}. In
some -- albeit unrealistic -- configurations however, HS states
can be perturbatively stable thanks to BPS conditions. In fact the
very existence of these BPS states is a peculiarity of String
Theory: no smooth supergravity solutions can  describe
supersymmetric states with spinning horizon in $D=4$
\cite{Cvetic:1995kv, Cvetic:1996kv, Peet:2000hn}. Precisely for
this reason their existence creates a puzzling situation insofar
as the microscopic counting of the degeneracy of states is matched
with the area of a stretched horizon, relying on Wald's
generalization of Bekenstein-Hawking entropy formula
\cite{Sen:2009bm, Dabholkar:2010rm}.

Aim of this note is to study properties of 1/2 BPS higher spin
states within the perturbative spectrum of heterotic string
compactifications on tori \cite{Narain:1986am} and (freely acting)
orbifolds
 \cite{Ferrara:1995yx}. With a bold abuse of
language one could dub these states as `small spinning
black-holes' \cite{Russo:1994ev, Sen:2009bm, Dabholkar:2010rm} in
that they have a finite microscopic `entropy', despite the lack of
a semi-classical supergravity description \cite{Elvang:2004rt,
Bena:2004de, Dabholkar:2006za, Bena:2007kg, Iizuka:2007sk}. As we
will see, their spin and large degeneracy characterize their
couplings to massless states, \eg gravitons and vector bosons. In
line with previous observations \cite{Chialva:2004xm,
Iengo:2006gm, Chialva:2009pf, Chialva:2009pg}, one should accept
to abandon a semi-classical description and determine the nature
of these peculiar string states in terms of observables such as
scattering amplitudes, decay rates and form factors. In this note,
relying on our previous work \cite{Bianchi:2010dy}, we will take a
further step in this direction. Contrary to recent analyses
\cite{Dudas:1999gz, Chialva:2005gt, Bianchi:2006nf, Lust:2008qc,
Anchordoqui:2009mm, Dong:2010jt, Feng:2010yx} of unstable string
excitations in phenomenologically viable scenari with TeV scale
tension and large extra dimension \cite{ArkaniHamed:1998rs,
Antoniadis:1998ig}, we will mostly focus on cases with at least
$\cN=2$ whereby BPS condition guarantee the perturbative stability
of higher spin states. One may hope that sectors with extended
supersymmetry could be embedded in realistic chiral models.

The plan of the paper is as follows. In Section \ref{BPSHS}, we
review the structure of $\cN =4$ 1/2 BPS multiplets with higher
spin. We identify HS multiplets of this kind in various heterotic
compactifications and discuss their vertex operators and BRST
conditions. We extend the analysis to non BPS HS states in Section
\ref{nonBPSHS}, where we display physical vertex operators for the
first two massive levels. We then study on-shell tri-linear
couplings of HS states in Section \ref{3-ptcoup}. Not
unexpectedly, we find that BPS states with arbitrary spin cannot
decay into lower spin states in perturbation theory. To study
their properties one should consider at least 4-point scattering
amplitudes: this is our task in Section \ref{4-ptamp}. We also
extract form factors and polarization effects at low energies.
Finally we conclude with a summary of our results, open issues and
puzzles in Section \ref{Final}.


\section{Higher Spin 1/2 $BPS$ $\cN=4$ multiplets
in $D=4$} \label{BPSHS}

Heterotic strings \cite{Gross:1985fr, Gross:1985rr} compactified
on six dimensional tori ${\bf T}^6$ enjoy $\cN=4$ supersymmetry in
$D=4$ \cite{Narain:1986am, Bianchi:1991eu, Chaudhuri:1995fk,
Bianchi:1997rf, Witten:1997bs, Angelantonj:1999xf,Bachas:2008jv}.
The spectrum includes super-multiplets with arbitrarily high spin.
Some of these multiplets saturate the BPS bound $M^2 = |\bf p_L|^2$,
where $p_L^i$ with $i=1,... 6$ denote the central charges the
gravi-photons couple to, and are shorter than generic long
multiplets. In the perturbative spectrum only 1/2 BPS and long
multiplets are present. Bosonic charged states in $1/2$ BPS
multiplets with maximal spin  correspond to the vertex
operators \beqn V^{(-1)}_{\cH_i} &=& \cH_{i,\mu_1 \cdots \mu_s}
e^{-\varphi} \psi^i e^{i{\bf p}_L {\bf X}_L} \bar\de X_R^{\mu_1}
\cdots\bar\de X_R^{\mu_s} e^{i{\bf p}_R {\bf X}_R} e^{i p X}
\label{eq vertex max spin 1}
\eeqn
with internal excitations in the ground state of the L-moving
sector, and \beqn V^{(-1)}_{\cH_\mu} &=& \cH_{\mu,\mu_1 \cdots
\mu_s} e^{-\varphi} \psi^\mu e^{i{\bf p}_L {\bf X}_L} \bar\de
X_R^{\mu_1} \cdots\bar\de X_R^{\mu_s} e^{i{\bf p}_R {\bf X}_R}
e^{i p X}
\label{eq vertex max spin 2}
\eeqn
with space-time excitations in the ground state of the L-moving sector. Clearly, at a given mass level there exist additional 1/2 BPS multiplets with lower spin.  \\
By construction, the tensors $\cH_{i,\mu_1... \mu_s}$ and
$\cH_{\mu,\mu_1... \mu_s}$ are totally symmetric in the
non-compact space-time indices $\mu_i$ and, in order for the
states to be BRST invariant, they should satisfy \beqn
\label{BRSBHcond1} p_L^i\cH_{i,\mu_1 \cdots
\mu_s}=p^{\mu_1}\cH_{i,\mu_1\mu_2 \cdots\mu_s}=\eta^{\mu_1\mu_2}\cH_{i,\mu_1\mu_2\mu_3 \cdots\mu_s}=0\\
p^\mu\cH_{\mu,\mu_1 \cdots \mu_s}=p^{\mu_1}\cH_{\mu,\mu_1\mu_2
\cdots\mu_s}=\eta^{\mu_1\mu_2}\cH_{\mu,\mu_1\mu_2\mu_3
\cdots\mu_s}=0 \,\,.\eeqn The tensor $\cH_{i,\mu_1... \mu_s}$
accounts for five\footnote{After imposing BRST conditions
(\ref{BRSBHcond1}) the internal index $i$ is allowed to run only
over the five directions orthogonal to the central charge vector
${\bf p}_{_L}$.} states of spin $s$, while $\cH_{\mu,\mu_1...
\mu_s}$ gives rise to spin $s+1$, $s$ and $s-1$ states according
to \beqn \label{YoungTab} \Yboxdim{10pt}
\yng(1)\otimes\yng(6)=\yng(7)\oplus\yng(6,1)\oplus\yng(5)
\,\,.\eeqn The reason why the second Young tableaux on the
right-hand side of (\ref{YoungTab}) corresponds to spin $s$ is
that, because of the BRST conditions (\ref{BRSBHcond1}), $\mu_i$
are effectively $SO(3)$ indices, thus
${\tiny\yng(1,1)\equiv\yng(1)}$. Therefore the number of bosonic
degrees of freedom is \beqn
n_B=2(s+1)+1+(5+1)(2s+1)+2(s-1)+1=(2s+1)8_B \,\,.\eeqn

Vertex operators for the fermionic states with maximal spin read
\beqn V^{(-1/2)}_{\Psi_{\a A}} &=&  \Psi_{\a A,\mu_1 \cdots \mu_s}
e^{-\varphi/2} S^\a \Sigma^A e^{i{\bf p}_L {\bf X}_L} \bar\de
X_R^{\mu_1} \cdots\bar\de X_R^{\mu_s} e^{i{\bf p}_R {\bf X}_R}
e^{i p X} \eeqn and \beqn V^{(-1/2)}_{\ov \Psi_{\dot\a A}} &=& \ov
\Psi^A_{\dot\a,\mu_1 \cdots \mu_s} e^{-\varphi/2} C^{\dot\a}
\Sigma^*_A e^{i{\bf p}_L {\bf X}_L} \bar\de X_R^{\mu_1}
\cdots\bar\de X_R^{\mu_s} e^{i{\bf p}_R {\bf X}_R} e^{i p X}
\,\,,\eeqn where $S^\a$, $C^{\dot\a}$ are $SO(1,3)$ spin fields
and $\Sigma^A$, $ \Sigma^*_A$ are $SO(6)$ spin fields. BRST
invariance requires \be p^\mu \bar\sigma_\mu^{\dot\a \a} \Psi_{\a
A,\mu_1 \cdots \mu_s} + p_{iL}\tau^i_{AB}\ov \Psi^{B\dot\a}_{\mu_1
\cdots \mu_s} = 0 \ee that allows to express $\ov
\Psi^{A\dot\a}_{\mu_1 \cdots \mu_s}$ in terms of $\Psi_{\a A,\mu_1
\cdots \mu_s}$ \be \ov \Psi^{A\dot\a}_{\mu_1 \cdots \mu_s} =
{1\over M^2} p^\mu \bar\sigma_\mu^{\dot\a \a}
p^i_{L}\bar\tau_i^{AB}\Psi_{\a B,\mu_1 \cdots \mu_s} \ee with
$M^2= -p\cdot p = |\bf p_L|^2$. Combing spin 1/2 from left-movers with
spin $s$ from Right-movers one gets $s+1/2$ and $s-1/2$. Taking
into account the degeneracy ${\bf 4}$ of $SO(5)\sim Sp(4) \subset
SO(6)$, the number of fermionic degrees of freedom turns out to be
\beqn n_F=4[2(s+1/2)+1]+4[2(s-1/2)+1]=(2s+1)8_F \eeqn Thus, these
multiplets contain $(2s+1)(8_B-8_F)$ complex charged states with
maximal spin $s_{hws}=s+1$.

An analogous analysis applies to higher spin 1/2 BPS
multiplets in $\cN=2$ compactifications such as (freely acting)
orbifolds. For details see appendix \ref{app higher spins in N=2}.

For some purposes, it is convenient to switch from the $\cN = 4$
notation in $D=4$ to an $\cN = (1,0)$ notation in $D=10$.

In $D=4$, the little group for massive particles is $SO(3)$, that
has rank one and admits only totally symmetric tensors as irreducible representations. Young Tableaux
have only one row. Equivalently working with $SO(4)\sim SU(2)_L
\times SU(2)_R$ one has un-dotted and dotted spinor indices which
are separately symmetrized. Two anti-symmetrized $SU(2)$ spinor
indices indeed represent the singlet.

In $D=10$ allowed Young Tableaux may have up to four rows since
$SO(9)$ has rank four and five anti-symmetrized vector indices are
equivalent to four. Yet in covariant notation, based on $SO(9,1)$
with rank five, it is necessary to consider anti-symmetric tensors
with up to 5 indices. Going from $D=10$ to $D=4$ another important
phenomenon takes place. $\cN = (1,0)$ super-multiplets in  $D=10$
may decompose into several $\cN = 4$ super-multiplets in $D=4$.
Indeed, even at the massless level, \ie setting $P=(p_\mu, 0)$,
the $\cN = (1,0)$ super-gravity multiplet in $D=10$ decomposes
into the $\cN = 4$ super-gravity multiplet along with six vector
super-multiplets in $D=4$. On the other hand, the $\cN = (1,0)$
massless vector super-multiplet in $D=10$ produces a single $\cN =
4$ massless vector super-multiplet in $D=4$.

Something similar happens for BPS and non-BPS massive multiplets.
Moreover, if some of the supersymmetries are broken upon
compactification, some of the states are eliminated or projected
out from the spectrum, and further decomposition of the
super-multiplets takes place. We will illustrate this point in Appendix \ref{app higher spins in N=2}.

\section{Massive non BPS super-multiplets}
\label{nonBPSHS}

For later purposes, let us also analyze non BPS states with
various spins in the higher mass levels of the perturbative
heterotic spectrum \cite{Gross:1985fr, Gross:1985rr,
Narain:1986am}. We relate the light-cone gauge counting, based on
$SO(8)$ for the left-moving superstring and $SO(24)$ for the
right-moving bosonic string, to the relevant little group
representations and show how BRST invariance allows to gauge away
the non-physical fields. We furthermore display the BRST invariant
vertex operators for the first and second mass levels explicitly.
Eventually one has to combine left- and right- movers imposing
level matching \cite{Narain:1986am}.

\subsection{The left-moving superstring sector}

Let us address the counting of degrees of freedom for the first
two massive levels of the superstring (in ten dimensions), and
show how to rearrange the states into representations of $SO(9)$,
the little group for massive states \cite{
Bianchi:2003wx, Beisert:2003te, Beisert:2004di,Hanany:2010da}. Before turning to
the massive levels, let us discuss the massless case. Due to its
simplicity, this level serves as a good example in the study BRST
invariance and identification of physical degrees of freedom.
Later on we will apply the lesson to the massive levels.

\subsubsection{Massless Level (Left Movers)}
In the light-cone gauge, the only bosonic states are
given by\footnote{With some abuse of notation we denote by $i,j,..
= 1,... 8$ the transverse directions.}
\begin{align}
\psi^{i}_{-1/2} | 0 \rangle \longrightarrow 8 &=\tiny{\yng(1)}
\end{align}
which correspond to a massless gauge boson in the open string
sector. In 10-dimensional covariant notation the state takes the
form
\begin{align}
A_{M}\psi^{M}_{-1/2} | 0 \rangle  \qquad \qquad  \cV_A(z) = A_{M}
\,e^{-\varphi}\, c\, \psi^{M} e^{ikX}
\end{align}
where the latter is the corresponding un-integrated vertex
operator in the $(-1)$ superghost picture that include the
$c$-ghost. The polarization vector $A_{M}$ is subject to the BRST
constraints, namely it has to satisfy $k^{M} A_{M} = 0$, $k^2 = 0$
which suggests that $A_{M}$ have nine independent components.
However, using the fact that the BRST operator is nilpotent, we
can add to the massless vertex operator the following operator
\begin{align}
\delta \cV(z) = \left[Q_{BRST} \, , {\cal U}(z)\right] \, ,
\end{align}
where $Q_{BRST}$ is the BRST charge given by
\begin{align}
Q_{BRST} = \oint {dz\over 2\pi i}\,\, e^{\varphi} \,\eta \,\psi_{M}\,
\partial X^{M} + ... \,\,.
\end{align}
For the massless level ${\cal U} (z)$ takes the form
\begin{align}
{\cal U} (z) =\Lambda \,\, e^{-2 \varphi} \,c\,\partial \xi\, e^{ikX}
\end{align}
with $k^2=0$. This allows to gauge away one un-physical component, thus a massless gauge boson in the left-moving sector
exhibits as expected 8 physical degrees of freedom in agreement with
 the light-cone gauge quantization.

\subsubsection{First Massive Level (Left Movers)}

At the first mass level, in the light-cone gauge, the bosonic
states are\footnote{Remember that, in the NS sector, states are
built out of the vacuum by the action of an odd number of
fermionic excitations. States with even fermionic excitation numbers are
eliminated by the GSO projection.}
\begin{align} \nonumber
\Yboxdim{10pt}
 \psi^i_{-3/2}|0\rangle\longrightarrow 8 &=\tiny{\yng(1)}\\ \nonumber
\Yboxdim{10pt}
\alpha_{-1}^{(i}\psi^{j)}_{-1/2}|0\rangle\longrightarrow
35&=\tiny{\yng(2)}\\ \label{eq SO(8) first massive level}
\Yboxdim{10pt}
\alpha_{-1}^{[i}\psi^{j]}_{-1/2}|0\rangle\longrightarrow
28&=\tiny{\yng(1,1)}\\ \nonumber \Yboxdim{10pt}
\psi^i_{-1/2}\psi^j_{-1/2}\psi^k_{-1/2} |0\rangle\longrightarrow
56&=\tiny{\yng(1,1,1)}\\ \nonumber \Yboxdim{10pt}
\delta_{ij}\,\,\alpha_{-1}^i\psi^j_{-1/2}|0\rangle\longrightarrow
1&= \bullet\,\,\,,
\end{align}
where as usual $(..)$ and $[..]$ denote symmetrization and
anti-symmetrization of the indices. Moreover, the symmetrized
states do not contain the trace, which is treated separately.

In 10-dimensional covariant notation the states are given by
\begin{align*}
&\widehat{E} \, \eta_{MN}\,   \alpha_{-1}^M\psi^N_{-1/2}|0\rangle
\hspace{15mm} \widehat{W}_M \,\psi^M_{-3/2}|0\rangle \hspace{15mm}
\widehat{B}_{[MN]} \, \alpha_{-1}^{[M}\psi^{N]}_{-1/2}|0\rangle \\
& \hspace{13mm}\widehat{H}_{(MN)} \,
\alpha_{-1}^{(M}\psi^{N)}_{-1/2}|0\rangle\hspace{15mm}
\widehat{C}_{[LMN]}\,\psi^L_{-1/2}\psi^M_{-1/2}\psi^N_{-1/2}
|0\rangle
\end{align*}
The un-integrated vertex operators corresponding to the above states
are given by
\begin{align}
\cV^{(-1)}_{\widehat{E}} &= \widehat{E} \,\, e^{-\varphi}  \, c\,\eta_{AB}  \,\partial X^A  \,\psi^{B} e^{ipX} \\
\cV^{(-1)}_{\widehat{W}} &= \widehat{W}_M \,\,    c\,e^{-\varphi}   \, \partial \psi^{M} \,e^{ipX}\\
\cV^{(-1)}_{\widehat{B}} &=
 \widehat{B}_{[MN]} \,\, e^{-\varphi }  \, c\,
 \partial X^{[M} \, \psi^{N]} e^{ipX} \\
\cV^{(-1)}_{\widehat{H}} &=
 \widehat{H}_{(MN)} \,\, e^{-\varphi }  \, c\,
 \partial X^{(M} \, \psi^{N)} \,e^{ipX} \\
\cV^{(-1)}_{\widehat{C}} &=
 \widehat{C}_{[MNP]} \,\, e^{-\phi }  \, c\,
\psi^{[M}\,\psi^{N} \,\psi^{P]}\, e^{ipX}\,\,.
\end{align}
Clearly, not all states correspond to physical degrees of freedom.
 BRST invariance allows one to gauge away the non-physical
components. As for the massless level before, one can add an
operator of the form
\begin{align}
\delta\cV = \left[Q_{BRST} \, , {\cal U} + {\cal U}'\right]\,\,,
\end{align}
where, in contrast to the massless case, we distinguish between
two different types of operators ${\cal U}$ and ${\cal U}'$. For the first massive level they take the form
\begin{align}
{\cal U} = e^{-2 \varphi} \,\partial \xi\, c\, \left[  \lambda_{[AB]}
\,\psi^{A} \,\psi^B + \omega_A \,\partial X^A \,\right] e^{ipX}
\qquad \text{and} \qquad {\cal U}' = e^{-2 \varphi} \,\partial^2 \xi
\, c\,\,\Lambda \,e^{ipX}\,\,,
\end{align}
with\footnote{Unless otherwise stated, we set $\ap =2$
henceforth.} $p^2=-2$. This implies that BRST invariance allows us
to gauge away one scalar, one vector and an anti-symmetric rank
two tensor. Let us be more precise and display the effect of
adding $\left[Q_{BRS}, {\cal U}  + {\cal U}' \right]$ to the polarization
tensors {\it viz.}
\begin{align*}
C_{[MNP]} &= \widehat{C}_{[MNP]} -i p_{[M} \lambda_{NP]} \\
H_{(MN)} &= \widehat{H}_{(MN)}-ip_{(M} \omega_{N)}\\
B_{[MN]} &= \widehat{B}_{[MN]}-\lambda_{[MN]} -i p_{[M} \omega_{N]}\\
W_M&=\widehat{W}_M-\omega_M -2i\Lambda p_M \\
E&= \widehat{E} +2\Lambda\,\,.
\end{align*}
As previously discussed, BRST invariance allows one to gauge away
$\widehat{B}_{[MN]}$, $\widehat{W}_M$, $\widehat{E}$. However, the
structure of the gauging is quite involved. One is left with a
totally anti-symmetric rank 3 tensor and a symmetric rank 2
tensor, which is in complete agreement with table \ref{table first
massive}. There we display the relation between the $SO(8)$
light-cone gauge group structure  and the little group $SO(9)$
structure. For the first massive level we see that the $SO(8)$
covariant states in equation \eqref{eq SO(8) first massive level}
can be rearranged into an  anti-symmetric rank 3 and a symmetric
rank 2 tensors invariant under the little group $SO(9)$.
\begin{table}[htbp]
\begin{center}
\begin{tabular}{cccc}
\hline
\hline
d.o.f & $SO(9)$ & & $SO(8)$\\
\hline
\\
$84$ & $\Yboxdim{8pt}\yng(1,1,1)$ &  & $\Yboxdim{8pt}\yng(1,1,1)+\yng(1,1)$\\
\\
$44$ & $\Yboxdim{8pt}\yng(2)$  & & $\bullet+\Yboxdim{8pt}\yng(1)+\yng(2)$\\
\\
\hline
\hline
\end{tabular}
\end{center}
\caption{Decomposition of $SO(9)$ into $SO(8)$ representations for
the first massive level. \label{table first massive}}
\end{table}

Finally, we display the un-integrated physical vertex operators. In the
$(-1)$-ghost picture they are given by \cite{Tanii:1987bk, Koh:1987hm,Dong:2010jt,
Feng:2010yx}
\begin{align}
\cV^{(-1)}_{C_{[MNP]}}=C_{[MNP]} \,\, e^{-\varphi} \,c\,\psi^{M}\,
\psi^{N}\,  \psi^{P}\,  e^{ipX}
\end{align}
and
\begin{align}
\cV^{(-1)}_{H_{(MN)}}=H_{(MN)} \,\, e^{-\varphi} \, c\,\partial
X^{(M}\,    \psi^{N)}\, e^{ipX}\,\,.
\end{align}
BRST invariance furthermore requires
\begin{align}
p^M\,C_{[MNL]}=0 \qquad p^M\,H_{(MN)}=0 \qquad \eta^{MN}\, H_{(MN)}=0  \,\,.
 \end{align}
in addition to the mass-shell condition $p^2 = - 2 = - M^2$.

\subsubsection{Second Massive Level (Left Movers)}

At the second massive level the number of states increases
drastically. Once again we display all states that arise from the
canonical quantization performed in the light-cone gauge\footnote{Note
that in contrast to the first massive level we do not display
every single state separately but rather in compact form. For
instance the state $\alpha_{-1}^i\psi^j_{-3/2}|0\rangle$ contains
a scalar, a symmetric and anti-symmetric rank 2 tensor.}
\begin{align*}
\Yboxdim{10pt}
\psi^i_{-5/2}|0\rangle\longrightarrow 8&=\tiny{\yng(1)}\\
\Yboxdim{10pt}
\alpha_{-1}^i\psi^j_{-3/2}|0\rangle\longrightarrow 64&=\bullet+\tiny{\yng(2)}+\tiny{\yng(1,1)}\\
\Yboxdim{10pt}
\alpha^i_{-2}\psi^j_{-1/2}|0\rangle\longrightarrow 64&=\bullet+\tiny{\yng(2)}+\tiny{\yng(1,1)}\\
\Yboxdim{10pt}
\psi^i_{-3/2}\psi^j_{-1/2}\psi^k_{-1/2}|0\rangle\longrightarrow 224&=\tiny {\yng(1)+\yng(1,1,1)+\yng(2,1)}\\
\Yboxdim{10pt}
\alpha_{-1}^i\alpha_{-1}^j\psi^k_{-1/2}|0\rangle\longrightarrow 288&=\tiny {\yng(1)+\yng(3)+\yng(2,1)}\\
\Yboxdim{10pt}
\alpha_{-1}^i\psi^j_{-1/2}\psi^k_{-1/2}\psi^l_{-1/2}|0\rangle\longrightarrow 448&=\tiny{\yng(1,1)+\yng(1,1,1,1)+\yng(2,1,1)}\\
\Yboxdim{10pt}
\psi^i_{-1/2}\psi^j_{-1/2}\psi^k_{-1/2}\psi^l_{-1/2}\psi^m_{-1/2}|0\rangle\longrightarrow 56&=\tiny{\yng(1,1,1,1,1)}
\end{align*}
In 10-dimensional covariant notation they take the form
\begin{align*}
\tiny{\yng(1,1,1,1,1)} & \qquad
\widehat{X}_{[MNPQR]}\,\,\psi^M_{-1/2}\psi^N_{-1/2}\psi^P_{-1/2}\psi^Q_{-1/2}\psi^R_{-1/2}|0\rangle \\
\tiny {\yng(2,1,1)}
&\qquad \widehat{Y}_{(M[N)PQ]}\,\, \alpha_{-1}^M\psi^N_{-1/2}\psi^P_{-1/2}\psi^Q_{-1/2}|0\rangle\\
\tiny{\yng(1,1,1,1)} & \qquad \widehat{A}_{[MNPQ]} \,\, \alpha_{-1}^M\psi^N_{-1/2}\psi^P_{-1/2}\psi^Q_{-1/2}|0\rangle\\
\tiny{\yng(3)} & \qquad  \widehat{Z}_{(MNP)}\,\, \alpha_{-1}^M \alpha_{-1}^N \psi^P_{-1/2}|0\rangle
\\
\tiny{\yng(1,1,1)} & \qquad  \widehat{B}_{[MNP]}\,\, \psi^M_{-1/2}\psi^N_{-1/2} \psi^P_{-3/2}|0\rangle
\\
\tiny{\yng(2,1)} & \qquad  \widehat{V}_{(M[N)P]}\,\, \psi^M_{-1/2}\psi^N_{-1/2} \psi^P_{-3/2}|0\rangle \qquad  \widehat{K}_{(M[N)P]}\,\, \alpha^M_{-1}\alpha^N_{-1} \psi^P_{-1/2}|0\rangle
\\
\tiny{\yng(2)} & \qquad  \widehat{M}_{(MN)} \alpha^{(M}_{-1} \psi^{N)}_{-3/2} |0\rangle \qquad
\widehat{L}_{(MN)} \alpha^{(M}_{-2} \psi^{N)}_{-1/2} |0\rangle\\
\tiny{\yng(1,1)} & \qquad  \widehat{D}_{[MN]} \, \eta_{KL}
\alpha_{-1}^L \psi^K_{-1/2}\psi^M_{-1/2}\psi^N_{-1/2}|0\rangle
\qquad
\widehat{F}_{[MN]} \alpha_{-1}^{[M}\psi^{N]}_{-3/2}|0\rangle \\
& \qquad  \widehat{G}_{[MN]} \alpha_{-2}^{[M}\psi^{N]}_{-1/2}|0\rangle
\\
\tiny{\yng(1)} & \qquad  \widehat{R}_M \, \eta_{AB} \psi_{-3/2}^A
\psi^B_{-1/2}\psi^M_{-1/2}|0\rangle \qquad
 \widehat{S}_M \, \eta_{AB} \alpha_{-1}^A \alpha_{-1}^B \psi^M_{-1/2}|0\rangle
\\
& \qquad  \widehat{T}_M \, \eta_{AB}  \psi_{-1/2}^A \alpha_{-1}^B
\alpha^M_{-1}|0\rangle  \qquad \widehat{U}_M
\psi^M_{-5/2}|0\rangle
\\
\bullet & \qquad \widehat{P}  \eta_{AB} \alpha_{-2}^A
\psi_{-1/2}^B |0\rangle \qquad  \widehat{E}  \eta_{AB}
\alpha_{-1}^A \psi_{-3/2}^B |0\rangle
\end{align*}
\begin{table}[htbp]
\begin{center}
\begin{tabular}{cccc}
\hline
\hline
d.o.f & $SO(9)$ & & $SO(8)$\\
\hline
\\
$126$ & $\Yboxdim{8pt}\yng(1,1,1,1,1)$ & & $ \Yboxdim{8pt} \yng(1,1,1,1,1)+\yng(1,1,1,1)$\\
\\
$156$ & $\Yboxdim{8pt}\yng(3)$ & & $\bullet+\Yboxdim{8pt}\yng(1)+\yng(2)+\yng(3)$\\
\\
$594$ & $\Yboxdim{8pt}\yng(2,1,1)$ & & $\Yboxdim{8pt}\yng(1,1)+\yng(1,1,1)+\yng(2,1)+\yng(2,1,1)$\\
\\
$231$ & $\Yboxdim{8pt}\yng(2,1)$ & & $\Yboxdim{8pt}\yng(1)+\yng(1,1)+\yng(2)+\yng(2,1)$\\
\\
$36$ & $\Yboxdim{8pt}\yng(1,1)$ & & $\Yboxdim{8pt}\yng(1)+\yng(1,1)$\\
\\
$9$ & $\Yboxdim{8pt}\yng(1)$ & & $\bullet+\Yboxdim{8pt}\yng(1)$\\
\\
\hline
\hline
\end{tabular}
\end{center}
\caption{Decomposition of $SO(9)$ into $SO(8)$ representations for
the second massive level. \label{Table second massive level} }
\end{table}

Clearly, not all of these represent physical degrees of freedom.
BRST invariance allows one to gauge away un-physical components.
Although we do not display the corresponding vertex operators one can shift them by
\begin{align}
\delta \cV = \left[{Q_{BRST}, {\cal U} + {\cal
U}'}\right]
\end{align}
with
\begin{align} \nonumber
 {\cal U}=
 e^{-2 \varphi } \,&\partial \xi \, c\,
 \Big[
  \alpha_{[MNPQ]} \,\,\psi^{M}\psi^{N}\psi^{P}\psi^{Q}  + \beta_{[MNP]} \,\, \psi^M \psi^N \partial X^P  + \gamma_{[M(N]P)} \,\,
  \psi^M \psi^{(N} \partial X^{P)} \\&+ \epsilon_{(MN)} \,\, \psi^{(M} \partial \psi^{N)}   +
  \kappa_{(MN)} \,\, \partial X^M \partial X^N + \lambda_{[MN]} \,\, \psi^{[M} \psi^{N]} + \theta_M \,\, \partial^2 X^M   \\ \nonumber &
  + \nu_M \,\, \psi^M \eta_{KL} \psi^K \partial X^{L} + \phi \,\,\eta_{AB} \partial X^A \partial X^{B} + \tau \,\, \eta_{AB} \psi^A \partial \psi^{B}
 \Big] e^{ipX}
\end{align}
and
\begin{align}
 {\cal U}' =
 e^{-2 \varphi } \,\partial^2 \xi \, c\,
 \Big[
\sigma_{[MN]}  \, \psi^{M} \, \psi^{N}  + \rho_{M} \partial X^{M}
 \Big] e^{ipX}
\end{align}
with $p^2=-4$ in both cases. This allows  to gauge away 2 scalars $\bullet$, 3
vectors ${\tiny \yng(1)}$, 2 symmetric rank 2 tensors ${\tiny
\yng(2)}$, 2 anti-symmetric rank 2 tensors ${\tiny \yng(1,1)}$, 1
hooked Yang diagram ${\tiny\yng(2,1)}$,  1 totally antisymmetric
rank 3 tensor ${\tiny\yng(1,1,1)}$, and 1 totally antisymmetric
rank 4 tensor ${\tiny \yng(1,1,1,1)}$. Thus the truly physical
degrees of freedom are represented by
\begin{align}
{\tiny \yng(1,1,1,1,1)} \qquad \qquad
{\tiny\yng(2,1,1) }\qquad \qquad  {\tiny\yng(3) }
\qquad \qquad {\tiny \yng(2,1) }\qquad \qquad {\tiny\yng(1,1)}
\qquad \qquad {\tiny\yng(1)}\,\,\,\,\,\,\,,
\label{eq degrees of freedom 2nd mass level}
\end{align}
which is again in complete agreement with the group theoretical
analysis. In Table  \ref{Table second massive level} we display
how the light-cone $SO(8)$ representatives can be rearranged into
representations of the little group $SO(9)$. We find exactly the
same degrees of freedom as displayed in equation \eqref{eq degrees
of freedom 2nd mass level}. Their physical vertex operators are
given by
\begin{align*}
\cV^{(-1)}_{X}&= X_{[MNPQR]}\,\, e^{-\varphi} \, c\,\psi^{M} \,\psi^N \,\psi^P\, \psi^Q\, \psi^{R} \,e^{ipX} \\
\cV^{(-1)}_{Y}&= Y_{(M[N)PQ]}\,\, e^{-\varphi}\,c\, \partial X^{M} \,\psi^N \,\psi^P \,\psi^{Q} \, e^{ipX} \\
\cV^{(-1)}_{Z}&= Z_{(MNP)} \,\,e^{-\varphi} \,c \,\partial X^{M} \,\partial X^{N} \,\psi^{P}  \, e^{ipX}\\
\cV^{(-1)}_U&= U_{(M[N)P]}\,\,e^{-\varphi} \,c \,\big[\partial \psi^M\, \psi^N \,\psi^P  -2 \,\partial X^M \,\partial X^N\, \psi^{P} \big]\, e^{ipX}\\
\cV^{(-1)}_V&= V_{[MN]}\,\,e^{-\varphi} \,c\,\left[ \psi^M \,\psi^N  \,\eta^{\perp}_{KL} \,\partial X^K \,\psi^L  -\frac{7}{2} \,\psi^M \,\partial^2 X^N -7\,\partial \psi^M \,\partial X^N \right] \,e^{ipX}\\
\cV^{(-1)}_W & = W_M \,\,e^{-\varphi}\,c\, \Big[ \partial X^M
\,\eta^{\perp}_{KL}\,  \psi^K \,\partial X^L -5 \,\psi^M
\,\eta^{\perp}_{KL}\,  \partial X^K\, \partial X^L +11 \,\psi^M\,
\eta^{\perp}_{KL} \, \psi^K \,\partial \psi^L\Big] \,e^{ipX}\,\,\,.
\end{align*}
Here $\eta^{\perp}_{MN}$ is the  $SO(9)$ invariant metric
\begin{align}
\eta^{\perp}_{MN}= \eta_{MN} - \frac{p^{M}\, p^N}{p^2} = \eta_{MN}
+\frac{p^{M}\, p^N}{M^2}\,\,.
\end{align}
In addition to the mass-shell condition $p^2=-4=-M^2$, BRST
invariance  implies the conditions
\begin{align}
p^M X_{[MNPQR]} = p^M Y_{(M[N)PQ]}= p^M Z_{(MNP)} = p^M
U_{(M[N)P]} = p^M V_{[MN]} = p^M W_M=0
 \end{align}
 and
 \begin{align}
 \eta^{MN}\, Y_{(M[N)PQ]} = \eta^{MN}\, Z_{(MNP)} = \eta^{MN} \,U_{(M[N)P]} =0\,\,.
\end{align}
which makes the polarization tensors manifestly $SO(9)$ covariant.

With significant more effort, one can determine the physical
vertex operators for bosonic states at higher mass levels. This is
beyond the scope of our present analysis. For a decomposition into
$SO(9)$ representations see \cite{Bianchi:2003wx, Beisert:2003te,
Beisert:2004di, Hanany:2010da}. Let us now turn to the
right-moving sector.

\subsection{The right-moving bosonic string sector}

The right-moving sector of the heterotic string consists of the
26-dimensional bosonic string \cite{Gross:1985fr, Gross:1985rr}.
In contrast to the superstring, there are only bosonic excitations
of the ground state. For simplicity and clarity, we will only
consider the Cartan generators associated to the 16 purely
right-moving directions. The generalization to the non-abelian
case, \ie including states with non-zero ${\bf p}_R$, is
straight-forward but slightly more involved since after
compactification their masses are moduli-dependent and may lead to gauge symmetry enhancement.

As for the superstring above, we will start with the massless case
and then turn to the massive levels.
\subsubsection{Massless level (Right movers)}
The only massless excitations for the bosonic sector\footnote{The
tachyonic state $|0\rangle$ which is also present in the bosonic
string does not  survive in the heterotic state,
due to the level matching condition.} are
\begin{align}
\ov \alpha^{a}_{-1} | 0 \rangle \longrightarrow 24 &=\tiny{\yng(1)}
\end{align}
where $a$ runs over the 24 dimensional light-cone coordinates and
describes a massless gauge boson in 26 dimensions. In
`covariant'\footnote{We put {\it covariant} in quotes since only
10 of the 26 R-moving bosonic coordinates have L-moving
counterparts.} (in $D=26$!!) notation the state and vertex
operator are given by
\begin{align}
B_A \ov \alpha^{A}_{-1} | 0 \rangle  \qquad \qquad \qquad \cV_B =  B_A
\,\,\ov c \,\ov \partial X^A \,e^{iK X}\,\,.
 \end{align}
Here $A$ runs over all 26 dimensions and BRST
invariance enforces $K_A B^A=0$ and  $K^2 = k_{10d}^2 + p_R^2 =0$.
Furthermore the vertex operator is unique up to the addition of an
operator of the form
 \begin{align}
\delta\cV = \left[ \ov Q_{BRST}\, ,  \ov {\cal U}\right]
 \end{align}
 where $\ov Q_{BRST}$ for the bosonic string takes the form
 \begin{align}
 \ov Q_{BRST}=\oint {d \ov z\over 2\pi i}  \, \,\frac{1}{2}\, \ov c \, \,\ov
 \partial X^{A} \,\ov \partial X_{A} (\ov z) + ...
 \end{align}
 and  for the massless level  $\ov {\cal U}$ is given by
 \begin{align}
\ov  {\cal U} = \Lambda \, e^{iKX}\,\,,
 \end{align}
which allows to gauge away one spurious state. Thus, taking into
account the BRST constraint $K_A B^A=0$, one has 24 pure physical
degrees of freedom, in complete agreement with the number one
obtains via light-cone gauge quantization.

\subsubsection{First massive level  (Right movers)}

At the first mass level, in the light-cone gauge, the states are
\begin{align}
\Yboxdim{10pt}
\delta_{ab}\,\ov \alpha^a_{-1}\ov \alpha^b_{-1}|0\rangle\longrightarrow 1&=\bullet\\
\Yboxdim{10pt}
\ov \alpha^a_{-2}|0\rangle\longrightarrow 24&=\tiny{\yng(1)}\\
\Yboxdim{10pt} \ov \alpha_{-1}^{(a} \ov
\alpha^{b)}_{-1}|0\rangle\longrightarrow 299&=\tiny{\yng(2)}\,\,.
\end{align}
In `covariant' 26-dimensional notation the states at the first mass level  are given by
\begin{align}
& \widehat{E}\,\, \eta_{AB}\,\ov \alpha^A_{-1} \ov
\alpha^B_{-1}|0\rangle \qquad \qquad \widehat{W}_A\,\,\ov
\alpha^A_{-2}|0\rangle \qquad \qquad \widehat{H}_{(AB)} \, \,\ov
\alpha_{-1}^{(A}\ov \alpha^{B)}_{-1}|0\rangle
\end{align}
and their corresponding vertex operators take the form
\begin{align*}
\cV_{\widehat{E}} &= \widehat{E} \,\,\ov c\, \eta_{AB} \, \ov
\partial X^A  \ov \partial X^{B} e^{ipX}\qquad \hspace{4mm}
\cV_{\widehat{W}} = \widehat{W}_A \,\,   \ov c\,  \ov
\partial^2 X^{A} e^{ipX}\\ & \hspace{20mm}V_{\widehat{H}}
=\widehat{H}_{(AB)} \, \, \ov c \, \ov  \partial X^{(A}  \ov
\partial X^{B)} e^{ipX}\,\,.
\end{align*}
As for the superstring before, not all components correspond to
physical degrees of freedom. BRST invariance again allows one
to gauge away the non-physical components. We can add an operator of
the form
\begin{align}
\delta\cV = \left[\ov Q_{BRST} \, , \ov {\cal U} +\ov {\cal U}' \right]
\end{align}
with
\begin{align}
\ov {\cal U} = \Gamma_A \,\,\ov \partial X^{A} e^{ipX} \qquad
\qquad \ov {\cal U}' =\Lambda \,\, \ov b\,\ov c\, e^{ipX}
\end{align}
where $p^2=-2$. Thus we can gauge away the massive scalar
$\widehat{E}$ and the massive vector $\widehat{W}_A$ and we are
left with a massive symmetric rank two tensor.  This symmetric
rank two tensor $H_{(AB)}$ accounts for the 324 degrees of freedom
obtained in the light-cone gauge quantization. In Table
\ref{bosonic string table first massive} we display the
decomposition of the SO(25) symmetric rank two tensor into the
$SO(24)$ representation. One sees that $H_{{AB}}$ indeed
contains one $SO(24)$ symmetric rank two tensor, one $SO(24)$
vector and one $SO(24)$ scalar.
\begin{table}[htbp]
\begin{center}
\begin{tabular}{cccc}
\hline
\hline
d.o.f & $SO(25)$ & & $SO(24)$\\
\hline
\\
$324$ & $\Yboxdim{8pt}\yng(2)$ &  & $\Yboxdim{8pt} \bullet+\yng(1)+\yng(2)$ \\
\\
\hline
\hline
\end{tabular}
\end{center}
\caption{Decomposition of $SO(25)$ into $SO(24)$ representations
for the first massive level. \label{bosonic string table first
massive}}
\end{table}

Below we display the vertex operator for the symmetric rank two tensor
\begin{align}
\cV_{H} (\ov z) = H_{(AB)} \,\,\ov c \,\ov \partial X^{A}  \,\ov
\partial X^{B} \,e^{ipX}  \,\,.
\end{align}
The BRST conditions read
\begin{align}
p^A \,\,  H_{(AB)} =0 \qquad \qquad  \eta^{AB} \,\,  H_{(AB)} =0
\end{align}
which reveals the $SO(25)$ invariance of the massive states
$H_{(AB)}$, in addition to $p^2 = -2 = -M^2$.

\subsubsection{Second mass level (Right movers)}

In light-cone gauge the second mass level contains the following
states
\begin{align}
\Yboxdim{10pt}
\ov \alpha^a_{-3}|0\rangle\longrightarrow 24&=\tiny{\yng(1)}\\
\Yboxdim{10pt}
\ov \alpha^a_{-1}\ov \alpha^b_{-2} |0\rangle \longrightarrow 576&=\tiny{\bullet + \yng(1,1) + \yng(2)}\\
\Yboxdim{10pt} \ov \alpha_{-1}^{a} \ov \alpha^{b}_{-1} \ov
\alpha_{-1}^{c}|0\rangle\longrightarrow 2600&=\tiny{\yng(1) +
\yng(3)}\,\,.
\end{align}
In `covariant' 26-dimensional form the relevant states are
\begin{align}
 \widehat{E}  \,\, \eta_{AB} \ov \alpha^A_{-1}\ov \alpha^B_{-2}
 &|0\rangle
 \qquad \qquad \hspace{12.5mm} \widehat{W}_A \ov \alpha^A_{-3}|0\rangle
 \qquad \qquad  \widehat{Y}_A \,\, \eta_{BC}\ov \alpha_{-1}^{A} \ov \alpha^{B}_{-1}
 \ov \alpha_{-1}^{C}|0\rangle\\
\widehat{B}_{[AB]} \,\, \ov \alpha^{[A}_{-1}\ov \alpha^{B]}_{-2}
&|0\rangle  \qquad \qquad \widehat{H}_{(AB)} \,\, \ov
\alpha^{(A}_{-1}\ov \alpha^{B)}_{-2} |0\rangle  \qquad \qquad
\widehat{S}_{(ABC)} \,\, \ov \alpha_{-1}^{A} \ov \alpha^{B}_{-1} \ov
\alpha_{-1}^{C}|0\rangle\,\,,
\end{align}
where  not all states correspond to pure physical degrees of freedom. Their corresponding vertex operators take the form
\begin{align*}
\cV_{\widehat{E}} &= \widehat{E}\,\, \ov c\, \eta_{AB} \,\ov \partial X^A\,  \ov \partial^2 X^B \,  e^{ipX} \\
\cV_{\widehat{W}}&= \widehat{W}_A \,\, \ov c\, \ov \partial^3 X^A \,  e^{ipX} \\
\cV_{\widehat{Y}} &= \widehat{Y}_A\,\, \ov c\, \eta_{BC} \,\ov
\partial X^A\,\ov \partial X^B\,  \ov \partial X^C \,  e^{ipX} \\
\cV_{\widehat{B}} &= \widehat{B}_{[AB]}\,\, \ov c\, \ov
\partial X^{[A}\,\ov \partial^2 X^{B]}\,    e^{ipX} \\
\cV_{\widehat{H}} &= \widehat{H}_{(AB)}\,\, \ov c\, \ov
\partial X^{(A}\,\ov \partial^2 X^{B)}\,    e^{ipX} \\
\cV_{\widehat{C}} &= \widehat{S}_{(ABC)}\,\, \ov c\, \ov \partial
X^{A}\,\ov \partial X^{B}\,\ov \partial X^{C}\,    e^{ipX}
\end{align*}
and analogously to the previous case BRST invariance allows us to
add operators of the type $\left[ Q_{BRST}, \ov {\cal U} +\ov  {\cal U}' \right]$
with
\begin{align}
\ov {\cal U}  &= \left[\Lambda\,\, \eta_{AB} \,\ov \partial X^A
\,\ov
\partial X^B  +  \alpha_A\,\, \ov \partial^2 X^A  + \beta_{(AB)}\, \,
\partial X^A \, \partial X^B\right] e^{ipX}\\
\ov {\cal U}' &= \gamma_{A} \,\, \ov b \, \ov c\, \ov \partial
X^A\, e^{ipX}
\end{align}
where $p^2=-4$.
\begin{table}[htbp]
\begin{center}
\begin{tabular}{cccc}
\hline
\hline
d.o.f & $SO(25)$ & & $SO(24)$\\
\hline
\\
$2900$ & $\Yboxdim{8pt}\yng(3)$ &  & $\Yboxdim{8pt} \bullet+\yng(1)+\yng(2) + \yng(3)$ \\
\\
$300$ & $\Yboxdim{8pt}\yng(1,1)$ &  & $\Yboxdim{8pt} \yng(1)+\yng(1,1) $ \\
\\
\hline
\hline
\end{tabular}
\end{center}
\caption{Decomposition of $SO(25)$ into $SO(24)$ representations for the second massive level.
\label{bosonic string table second massive}}
\end{table}
Thus the truly physical degrees of freedom are just the
antisymmetric rank 2 tensor and the completely symmetric rank 3
tensor.

 In Table \ref{bosonic string table second massive} we
display how the $SO(25)$ massive states are decomposed into
$SO(24)$ light-cone representations. Note that the antisymmetric
rank 2 and the completely symmetric rank 3 tensors account for the
states obtained via the light-cone gauge quantization. Below we
display the vertex operators
\begin{align}
\cV_{B} &= B_{[AB]} \,\, \ov c \, \ov \partial X^{[A} \, \ov
\partial^2 X^{B]} \, e^{ipX} \qquad \qquad \cV_{S} = S_{(ABC)} \,\,
\ov c \, \ov \partial X^{A} \, \ov \partial X^{B} \, \ov \partial
X^{C}\, e^{ipX}
\end{align}
which have to satisfy the mass-shell condition $p^2 = - 4 = -M^2$
and the BRST constraints
\begin{align}
p^A B_{[AB]} = p^A S_{(ABC)} =0 \qquad \qquad \eta^{AB} \,
S_{(ABC)} =0\,\,.
\end{align}
The latter makes the polarization tensors manifestly $SO(25)$ covariant.

\subsection{Heterotic string, Type I and Type II superstring}

Though straightforward, proceeding to higher levels becomes more
and more cumbersome, except possibly for the highest spin states
(\ie the leading Regge trajectory of the graviton). In a recent
paper \cite{Hanany:2010da} (see also \cite{Bianchi:2003wx,
Beisert:2003te, Beisert:2004di}), the massive spectrum of
Heterotic, Type II, and Type I superstrings has been assembled
into $SO(9)$ representations. On the other hand, production of
open string Regge resonances in Type I and related models with
open and unoriented strings has been studied in
\cite{Dudas:1999gz, Chialva:2005gt, Bianchi:2006nf, Lust:2008qc,
Anchordoqui:2009mm, Dong:2010jt, Feng:2010yx}. Given the expected
duality between Heterotic and Type I strings, one may wonder if
there is any way to compare HS states in the two descriptions.
Clearly non BPS states are unstable and can decay into lower spin
states \cite{Chialva:2009pf, Chialva:2009pg}. Yet, 1/2 BPS states
should match on the two sides. The Type I counterpart of 1/2 BPS
HS states should be HS excitations of wrapped D-strings. Indeed
Heterotic / Type I duality has been carefully tested at the level
of the BPS spectra both in toroidal compactifications with or
without tensor structure and in other pairs (\eg freely acting
orbifolds etc). The relevant helicity super-trace (`topological
index' \cite{Kiritsis:2007zz}) does not allow to identify the spin
unambiguously as we will momentarily see. Despite this, one can
explore dynamical properties of perturbatively stable HS in Type I
or Type II theories using methods similar to those exploited in
\cite{D'Appollonio:2010ae}.

\section{Tri-linear couplings of charged higher spin BH's}
\label{3-ptcoup}

Let us now discuss tri-linear couplings of higher spin
states\footnote{For recent work on this issue from a different vantage
point, see \eg \cite{Sagnotti:2010at, Fotopoulos:2010ay}.}.

We will first consider 1/2 BPS states. In the NS sector vertex
operators are of the form \be V^{(-1)}_{1/2 BPS} = A_M \psi^M
e^{-\varphi} e^{i PX} V_R \ee where $\psi^M = (\psi^\mu, \psi^i)$
and $V_R$ accounts for the Right-movers.

\subsection{Left-Movers (Superstring)}

Let us focus on the Left-moving part first. BRST invariance
imposes \be P^2 = 0 = p^2 + |\bf p_L|^2 \qquad P^M A_M = p^\mu a_\mu +
p^i_L \Phi_i = 0 \ee \ie the Left-moving part is exactly identical
to the Left-moving part of a massless vertex in $D=10$.
(Super)ghost charge violation requires to consider \be
G_L(z_1,z_2,z_3)= \langle cV^{(-1)}_L(z_1) cV^{(0)}_L(z_2)
cV^{(-1)}_L(z_3)\rangle \ee where \be V^{(0)}_L(z) = A_M (\de X^M
+ i P\psi \psi^M)e^{i PX} \ee Performing the contractions one gets
\be G_L(z_1,z_2,z_3) = [A_1(P_2-P_3) A_2A_3 + A_2(P_3-P_1) A_3A_1
+ A_3(P_1-P_2) A_1A_2] \delta(P_1 + P_2 + P_3)  \ee that is
independent of $z_i$, totally anti-symmetric and vanishing
on-shell for real momenta.

As a result there is no physical (decay) process involving three
1/2 BPS states at tree level! This is largely a consequence of the
kinematics and the absence of quantum corrections to the mass and
charge and is independent of the spin of the states which is mostly
contributed by the right-moving bosonic string excitations. We
thus expect this non-renormalization property to hold to all
orders in perturbation theory and even non-perturbatively. This
seems to raise a puzzle in the microscopic counting of states that
reproduce the degeneracy of small 1/2 BPS Black Holes
\cite{Sen:2009bm, Dabholkar:2010rm}. Indeed (extended)
supersymmetry in $D=4$ is only compatible with zero horizon
rotation (horizon spin), spin can only emerge from hair d.o.f. \ie
from the global supersymmetry parameter that carries spinor
indices and vanishes at the horizon (susy enhancement!) while
assuming constant value at infinity. HS (BPS) states are
a peculiarity of String Theory and may require some
reconsideration. While it is important to study the
(thermo)dynamical properties of these HS states, one should keep
in mind that by going from weak coupling ($g_s << 1$) where the
(heterotic) string description is reliable to strong coupling
($g_s >> 1$) where the supergravity description should take over,
non-perturbative effects may `replace' HS supermultiplets with
minimal spin ones \ie vector multiplets. Indeed, the only
`protected' information is coded in the helicity supertace
\cite{Kiritsis:2007zz} \be \cB_4 = Tr(-)^{2h} (2h)^4 \ee
and one can easily check that \be \cB^{1/2 BPS}_{4, S} = (2S+1)
\cB^{1/2 BPS}_{4,S=0} \ee where $S$ is the `intermediate' spin in
the super-multiplet with $(2S+1) (8_B + 8_F)$ d.o.f. so that $S=0$
correspond to the `standard'(massless or 1/2 BPS) vector multiplet. Yet in perturbation theory we don't see any trace of
the `instability' of these HS states ...

The situation drastically changes if one of the three states (say
the one at $z_2$) is non BPS. In order to have a non-vanishing
coupling with two 1/2 BPS states one should select the non-BPS
state properly. For instance at the first massive level in the NS
sector one has two kinds of physical vertex operators \be V^{(0)}_{H,L} =
H_{MN} [\de X^{M}\de X^{N} + {i\over 2} P\psi (\psi^{M}\de X^{M}+
\psi^{N}\de X^{M}) + {1\over 2} (\psi^{M}\de \psi^{M}+ \psi^{N}\de
\psi^{M}) ] e^{i PX}\ee and \be V_C = C_{LMN} (\de X^{L} +i P\psi
\psi^{L}) \psi^{M}\psi^{N} e^{i PX}\,\,.\ee BRST symmetry implies \be
H_{MN} = H_{MN} \qquad \eta^{MN} H_{MN} = 0 \qquad P^M H_{MN} = 0
\ee leading to a massive spin 2 particle with 44 physical
polarizations and \be C_{LMN} = - C_{MLN} = - C_{LNM} = - C_{NML}
\qquad P^M C_{LMN} = 0 \ee leading to a massive 3-form with 84
physical polarizations.

The latter couples to two 1/2 BPS states through the symmetric
combination \be (P_1-P_2)^L A_1^M A_2^N C_{LMN} \ee which is gauge
invariant under $A_i\rightarrow A_i + \xi P_i$ thanks to momentum
conservation and transversality of $C_{LMN}$.

Similarly, the non-vanishing gauge-invariant coupling to the spin
2 combination reads \be H_{MN} F_1^{ML} F^N_{2L} \,\,,\ee where \be
F_{MN} = P_M A_N - P_N A_M \ee  thus exposing the manifest gauge
invariance wrt $A_i\rightarrow A_i + \xi P_i$.

In order to determine the on-shell coupling of two bosonic 1/2 BPS
vertex operators to higher mass non BPS states it proves
convenient to start from the OPE of two fermionic vertex
operators. The latter can be both taken to be in the canonical
(-1/2) picture and produce bosons in the canonical (-1) picture.
Expanding wrt the middle point \cite{Fotopoulos:2010cm} allows one
to extract the on-shell coupling to higher mass states. ${\cal
N}=4$ supersymmetry transformations determine the coupling between
two bosonic states. The relevant OPE of two fermionic 1/2 BPS
vertex operators reads
\begin{align} \nn e^{-\varphi/2} S^{A_1} &e^{iP_1X}(z_1)
e^{-\varphi/2} S^{A_2} e^{iP_2X}(z_2)
\approx  \\
&\quad z_{12}^{P_1P_2 - 1} \,\,e^{-\varphi} \,(1 + ...)\,\,
e^{i(P_1+P_2)X} \label{eq OPE fermions} (1 + z_{12} (P_1-P_2)\de
X...)
\\& \hspace{10mm} \times (\Gamma_M^{A_1A_2} \psi^M + z_{12}
\Gamma_{MNL}^{A_1A_2} \psi^M \psi^N\psi^L + z^2_{12}
\Gamma_{M_1...M_5}^{A_1A_2} \psi^{M_1}... \psi^{M_5}+ ...)
\nn\end{align} where dots stand for derivatives.

Let us take a closer look at the coupling of two BPS fermions to
the first massive excitations, $C_{[LMN]}$ and $H_{(MN)}$. The
antisymmetric rank $3$ tensor $C_{[LMN]}$ will couple through
$\Gamma^{A_1A_2}_{MNL}$ in \eqref{eq OPE fermions}. On the other
hand $H_{(MN)}$ will couple through
$(P_1-P_2)_{(N}\Gamma_{M)}^{A_1A_2}$. Note that in contrast to the
coupling of three 1/2 BPS states the coupling of two 1/2 BPS
states to a non-BPS state at the first mass level is  (graded)
symmetric. At next level one has a coupling through $\Gamma^{A_1A_2}_{M_1...M_5}$ which is antisymmetric.

Generalizing this statement to higher mass levels, one observes
that the coupling of a massive non-BPS vertex operator at odd
(even) level to two 1/2 BPS vertex operators is (anti) symmetric
under the exchange of the latter two.

\subsection{Right-Movers (Bosonic String)}

In the Heterotic string one has to consider the contribution of
the right-moving sector to the 3-point amplitudes. The relevant
correlation function on the world-sheet is of the form \be
G_R(\bar{z}_1, \bar{z}_2, \bar{z}_3) = \langle \ov c H_1[\bar\de^{k_i}
X_R^{A_i}]e^{iP_{1R}X_R}(\bar{z}_1) \ov c H_2[\bar\de^{m_j} X_R^{B_j}]
e^{iP_{2R}X_R}(\bar{z}_2) \ov c H_3[\bar\de^{n_k} X_R^{C_k}]
e^{iP_{3R}X_R}(\bar{z}_3) \rangle \ee
where
\be H[\bar\de^{k_i}
X_R^{A_i}]=
H_{A_1...A_n}\bar\de^{k_1}X^{A_1}...\bar\de^{k_n}X^{A_n}=
H_{A_1...A_n}\left[{\de\over \de \beta_{A_1}^{(k_1)}} ...{\de\over
\de \beta_{A_n}^{(k_n)}}\exp\sum_k
\beta_{A}^{(k)}\bar\de^{k}X^{A}\right]_{\beta_{A}^{(k)}=0} \ee
so that
\be G_R(\bar{z}_1, \bar{z}_2, \bar{z}_3) =
\ov c(\ov z_1)\ov c(\ov z_2) \ov c(\ov z_3)
H^1_{...A_i...}...{\de\over \de \beta_{A_i}^{(k_i)}}...
H^2_{...B_j...}...{\de\over \de \beta_{B_j}^{(m_j)}}...
H^3_{...C_k...}...{\de\over \de
\beta_{C_k}^{(n_k)}}...\cW|_{\beta_{A}^{(k)}=0} \ee where \be \cW
= \prod_{i<j} \bar{z}_{ij}^{P_iPj} \exp{i \sum_{i\neq j} (-)^{k_j}
k_j! {P_i\beta^{(k_j)}_j\over \bar{z}_{ij}^{k_j+1}}} \exp{
\sum_{i< j} (-)^{l_i + k_j} (l_i+k_j)!
{\beta^{(l_i)}_i\beta^{(k_j)}_j\over \bar{z}_{ij}^{k_j+l_i}}}\,\,. \ee

Except for the first Regge trajectory involving only $\bar\de
X^{A}$ \ie  all $k_n=1$, the result is unwieldy. Consider three HS
states with spin $S_i$, thus each containing $S_i$ conformal
fields $\bar\de X^{A}$ in the vertex operator. Provided
$\sum_i(S_i - T_i)= 2K$ (even number) with $T_i\le S_i$, $T_i$
indices can contract with the momenta $P_j$ of the other
insertions. The remaining ones should contract with one another.
Denoting $S_i'= S_i - T_i$ then $S'_{12}= (S'_1+S'_2-S'_3)/2$
indices of $H_1$ will contract with $H_2$ and so on.

For totally symmetric tensors  -- not necessarily with maximal
spin $S_R = N_R$, as in the first Regge trajectory -- the relevant
tri-linear coupling reads \bea \sum_{T_i\le S_i: \sum_i(S_i -
T_i)= 2K} (\ap)^{T_1+T_2+T_3/2}\prod_i\left(^{S_i}_{T_i}\right)
\left(^{S'_i}_{S'_{i,i+1}}\right) P_{23}^{A_1} ...
P_{23}^{A_{T_1}} H^1_{A_1...A_{T_1}}{}^{D_{1} ...
D_{S'_{31}}} {}_{F_{1} ... F_{S'_{12}}}  \nn\\
P_{31}^{B_1} ... P_{31}^{B_{T_2}} H^2_{B_1...B_{T_2}}{}^{F_{1} ...
F_{S'_{12}}} {}_{E_{1} ... E_{S'_{23}}} P_{12}^{C_1} ...
P_{12}^{C_{T_3}} H^3_{C_1...C_{T_3}}{}^{E_{1} ...
E_{S'_{23}}}{}_{D_{1} ... D_{S'_{31}}} \eea

When at least two vertex operators are identical (say 1 and 3),
the result is (anti)symmetric under the exchange of the two
depending of whether $S_2$ is even (or odd).

In the special case in which some non-abelian current algebra
survives, at lowest level ($N_R = 1$) one has the currents $J^a$. At
the next level $N_R = 2$ one finds the primary $H_a = d_{abc} J^b
J^c$ where $d_{abc} = Tr (T_a\{T_b, T_c\})$ is totally symmetric
and traceless wrt the Cartan-Killing metric in the absence of
abelian ideals.

The resulting coupling to two currents is \be \langle J^a(z)
 H^b(u) J^c(w)  \rangle = { d^{abc} \over (z-u)^2 (w-u)^2} \ee which
is manifestly symmetric as expected.

It is amusing that a similar situation prevails in Type I and
other models with open and unoriented strings even for non BPS HS
states \cite{Dudas:1999gz, Chialva:2005gt, Bianchi:2006nf,
Lust:2008qc, Anchordoqui:2009mm, Dong:2010jt, Feng:2010yx}. Indeed
twist symmetry implies that levels differing by an odd integer
have opposite symmetry under the exchange of the two ends of the
(open unoriented) string.

\section{Scattering Amplitudes, Form Factors and all that}
\label{4-ptamp}

In order to further explore the dynamical properties of BPS HS
states, we will compute the 4-point scattering amplitudes
involving states of this kind at tree level \cite{Bianchi:2010dy}.
Due to central charge conservation we will need at least two such
states in the process. It should also be clear that, thanks to
mass and charge conservation, there is no physical decay amplitude
of one massive BPS state to three (or more) states be either BPS
or not. Indeed $M_{BPS}=|\bf p_L| = |\sum_\ell \bf p_L(\ell)| \le
\sum_\ell |\bf p_L(\ell)| \le \sum_\ell M(\ell)$ and the process is
kinematically allowed only when all momenta, including the
internal components representing central charges, are aligned so
that there is no true scattering.

For simplicity, we will focus on the 4-point amplitude containing two
massless gauge bosons of the visible sector\footnote{Following \cite{Bianchi:2010dy}  the visible sector denotes the gauge sector already present in 10 dimensions.} and two higher spin fields that describe for
instance the scattering of a `photon' off a very massive HS
object. We will analyze the pole structure and low energy limit
and identify the exchanged particles. This will allow us to
determine the coupling of such higher spin states to the massless
particles such as the graviton and dilaton. Along the way we need
the generating function for a correlator containing an arbitrary
number of the conformal fields $\partial X^{A}$, whose derivation
is presented in appendix \ref{app correlator}.

More specifically, we are interested in the 4-point amplitude
\begin{align}
\cA_{AA\rightarrow \cH_i\bar{\cal H}_j} = \Big\langle V^{(-1)}_{\cH_i} (z_1) \, V^{(0)}_{A}(z_2) \, V^{(0)}_{A}(z_3)\, V^{(-1)}_{\cH_j} (z_4) \Big\rangle
\label{eq 4 point amplitude internal}
\end{align}
with the vertex operators for the gauge fields and the higher spin fields
\begin{align}
 V^{(-1)}_{\cH_i} &= \cH_{i,\mu_1 \cdots
\mu_s} e^{-\varphi} \psi^i e^{i{\bf p}_L X} \bar\de
X^{\mu_1} \cdots\bar\de X^{\mu_s} e^{i{\bf p}_R X} e^{i p
X} \\
V^{(0)}_{A} &= a_{\mu}\left(\partial
X^\mu-ik \cdot \psi\psi^\mu\right)\bar{J}^a
e^{ik X}\,\,.
\end{align}
As discussed in section \ref{BPSHS}, $\cH_{i}$ accounts for 5 spin
$s$ states. The BRST conditions for the projections read
\begin{align}
k^{\mu} a_{\mu} =p^{\mu_i}\, \cH_{i,\mu_1 \cdots\mu_s}  =p^{i}_L\, \cH_{i,\mu_1 \cdots\mu_s} =\eta^{\mu_i \mu_j}  \, \cH_{i,\mu_1 \cdots\mu_s}=0 \qquad \qquad  \forall i,j \neq 0
\end{align}
and the mass is given by
\begin{align}
M^2=|{\bf p}|^2_L=|{\bf p}|^2_R +2 (s-1)\,\,.
\end{align}
As for the three-point functions discussed in the previous
section, the amplitude splits into a holomorphic and an
anti-holomorphic part. Decomposing the HS field $H_{i\mu_1...\mu_s}$ into the tensor product of an internal vector $v_i$ and
a space-time tensor $h_{\mu_1...\mu_s}$
\begin{align*}
 H_{i\mu_1...\mu_s}=v_{i} \otimes h_{\mu_1...\mu_s}
\end{align*}
the left-moving part yields
\cite{Bianchi:2010dy}
\begin{align}
&\cW^L(z_i) = {v_1}_{i} \,  {a_{2}}_{\kappa}\,  {a_{3}}_{\lambda} \, {v_4}_{j}\,\, \Big\langle e^{-\phi(z_1)} \,\, e^{-\phi(z_4)}\Big\rangle \Big \langle
e^{i {\bf p}_L {\bf X}_L}(z_1)
\,e^{-i{\bf p}_L {\bf X}_L}(z_4)\Big\rangle \Big \langle
\psi^{i}(z_1) \psi^{j}(z_4)\Big\rangle\,\,
\label{eq left-moving part 4 point st}
\\ \nn
& \hspace{3mm}\Big\langle \psi^{\mu_0}\,e^{ip_1 X}(z_1) \left[\partial X^{\kappa} - i (k_2\cdot\psi)
\psi^{\kappa}\right]e^{ik_2 X}(z_2) \left[\partial X^{\lambda} - i (k_3\cdot\psi\right]\psi^{\lambda}
)e^{ik_3X}(z_3) \psi^{\nu_0}\,e^{ip_4{X}}(z_4)\Big\rangle\,\,.
\end{align}
The right-moving part takes the form
\begin{align} \label{eq right-moving part 4-point st}
\cW^R (\bar z_i) &= {h_1}_{\mu_1...\mu_s } {h_4}_{\nu_1...\nu_s }
\Big\langle \bar{J}^{a_2}(\bar{z}_2)\bar{J}^{a_3}(\bar{z}_3) \Big\rangle \,\, \Big \langle \ e^{i{\bf p}^1_R {\bf X}_R}(\bar{z}_1) e^{i{\bf p}^4_R {\bf X}_R}(\bar{z}_4) \Big\rangle
\\ \nn
& \hspace{4mm}\Big\langle\bar\de
X^{\mu_1} \cdots\bar\de X^{\mu_s}  e^{i p_1
X} (\bar z_1) \, e^{i k_2
X} (\bar z_2) \, e^{i k_3
X} (\bar z_3) \,
\bar\de
X^{\nu_1} \cdots\bar\de X^{\nu_s}  e^{i p_4
X} (\bar z_4)\Big\rangle\,\,.
\end{align}

The details of the calculation are relegated  to appendix \ref{app
details of amplitude}. After reinstating normalization factors,
one finds
\begin{align}
\label{eq final result} {\cal A} =-  {g_{_{YM}}^2\over M_s^2}
(2\pi)^{4}\delta(\Sigma_ip_i) \frac{1}{k_2k_3} v_1 \cdot v_4 \,
(p_4 f_2 f_3 p_4)     {h_1}_{\mu_1...\mu_s }\,
{h_4}_{\nu_1...\nu_s } \, {\cal F}_{SV}\,{\cal
J}^{\mu_1...\mu_s\nu_1...\nu_s}_R
\end{align}
where ${\cal J}_R $ is given by equation \eqref{eq final equation
for J} in appendix \ref{app details of amplitude}, $f_i$ denotes
the linearized field strength
\begin{align}
{f_i}_{\mu\nu} = {k_i}_{\mu} {a_i}_{\nu}-  {a_i}_{\mu} {k_i}_{\nu}
\end{align}
the contraction $(p_4 f_2 f_3 p_4)$ is understood as
\begin{align}
{p_4}_{\mu} \,{f_2}^{\mu\nu}\, {f_3}_{\nu\rho} \,{p_4}^{\rho}
\end{align}
and the Shapiro-Virasoro ${\cal F}_{SV}$ factor is given by (recall $\alpha'=2$)
\begin{align}
\label{eq S-V factor}
{\cal F}_{SV} =
\frac{\Gamma(1+k_2\,k_3)\Gamma(k_3\,p_4)\Gamma(k_3\,p_1)}
{\Gamma(2-k_2\,k_3)\Gamma(-k_3\,p_4)\Gamma(-k_3\,p_1)}
 =  \frac{\Gamma \left( 1-\frac{t}{2} \right)
 \Gamma\left( \frac{M^2}{2} -\frac{s}{2} \right)\Gamma\left( \frac{M^2}{2}-\frac{u}{2}
\right)} {\Gamma \left( 2+\frac{t}{2} \right) \Gamma\left(
\frac{s}{2} -\frac{M^2}{2}\right)\Gamma\left(\frac{u}{2} -\frac{M^2}{2} \right)}\,\,,
\end{align}
where in the last step we used the usual definitions of Mandelstam
variables
\begin{align}
s&=-(k_2 +p_1)^2 = -2 k_2p_1 + M^2 \\  t&=-(k_2 +k_3)^2 =
-2 k_2 k_3 \\  u&=-(k_3 +p_1)^2 = -2 k_3p_1 + M^2\,\,.
 \end{align}
Clearly, the amplitude is invariant under the gauge transformation
\begin{align}
{a_r}_{\mu} \longrightarrow  {a_r}_{\mu} + i \Lambda_r k_{\mu}
\end{align}
but also under a  fake ``gauge transformation''
\begin{align}
{v_\ell}_{M} \longrightarrow  {v_\ell}_{M} + i \Lambda_\ell
{P_\ell}_{M} \, . \label{eq gauge transfo}
\end{align}
Here $v^M= (0,v^i)$ is the ten-dimensional left-handed
polarization vector and $P_M$ denotes the whole ten-dimensional
momentum. To expose this invariance  let us perform such a
transformation for $v_1$. The product $v_1 \cdot v_4$ gives
${\bf p}^1_L \cdot v_4$, since  $v^M_4$ vanishes in  space-time. Due to central charge
conservation ${\bf p}^1_L =-{\bf p}^4_L$ and with BRST
invariance  the term ${\bf p}^1_L \cdot v_4$ vanishes. Analogously
one can show gauge invariance of the amplitude under a
transformation \eqref{eq gauge transfo} for $v_4$. The reason for
this ``gauge'' invariance of $v_i$ lies in the fact that the
left-moving  part of a BPS state vertex operator is exactly the
one of a massless 10-dimensional gauge boson vertex operator.
Although we will not exploit this fake ``gauge invariance'' any
further in the present analysis, we would like to add that it
might prove useful in the computation of amplitudes involving several BPS
states.

Using momentum conservation and BRST invariance allows us to write
the amplitude in a way that displays its symmetries, which are
manifest in the form
\begin{align}
{\cal A} =-   {g_{_{YM}}^2\over M_s^2}
(2\pi)^{4}\delta(\Sigma_ip_i)\frac{ v_1 \cdot v_4 }{k_2k_3}  \, & \left[\frac{1}{2} p_1 (f_2 f_3 +f_3f_2) p_4 +\frac{1}{4} k_2 \cdot k_3 (f_2 \cdot f_3)  \right] \\ & \hspace{20mm}
\times {h_1}_{\mu_1...\mu_s }\, {h_4}_{\nu_1...\nu_s }\, {\cal F}_{SV}\,  {\cal J}^{\mu_1...\mu_s\nu_1...\nu_s}_R\, \nn
\end{align}
where $ (f_2 \cdot f_3) = {f_2}_{\mu\nu} {f_3}^{\mu \nu}$. With $
{\cal J}_R$, see eq. \eqref{eq final equation
for J}, being symmetric under the exchange of $1
\leftrightarrow 4$ as well as $2 \leftrightarrow 3$ in this from
the amplitude reveals the expected symmetries under the exchange
of  $1 \leftrightarrow 4$ and $2 \leftrightarrow 3$.

\subsection{Specific case $s=2$}

In order to further simplify the discussion, we will henceforth
consider the specific case of $s=2$. In this case one can display
$\cJ_R$ explicitly and compactly. Then the coupling of such a spin
$2$ state to massless particles such as graviton, dilaton or
antisymmetric rank two tensor can be extracted by looking at the
residue of the $t$-channel pole ${1}/{k_2k_3}$.

The amplitude for $s=2$ takes the form
\begin{align} \nn
{\cal A} =- {g_{_{YM}}^2\over M_s^2}&
(2\pi)^{4}\delta(\Sigma_ip_i) \frac{1}{k_2k_3}  \,\, {\cal
F}_{SV}\,\,v_1 \cdot v_4 \, (p_4 f_2 f_3 p_4)  \times \left[ L_0 +
\frac{k_3\,p_4 +1}{k_3p_1} L_1 + \frac{k_3\,p_1 +1}{k_3p_4} L_2
\right. \nn \\  &\left.
 +\frac{(k_3 p_4+2)(k_3p_4+1)}{(k_3 p_1)(k_3 p_1 -1) } L_3 + \frac{(k_3 p_1+2)(k_3p_1+1)}{(k_3 p_4)(k_3 p_4 -1) } L_4 \right]
\label{eq final result s=2}
\end{align}
with the $L_i$'s given by
\begin{align*}
L_0 &=2 \,({h_1} \cdot {h_4} ) - 4 (k_2\, h_1\, h_4\, k_2) - 4 (k_3\, h_1\, h_4\, k_3)\\&\hspace{3mm} + (k_2
h_1 k_2) (k_2 h_4 k_2) + (k_3
h_1 k_3) (k_3 h_4 k_3) +
 4 (k_2 h_1 k_3) (k_2 h_4 k_3) \\
L_1 &= 4 (k_3 h_1 h_4 k_2) - 2 (k_2 h_1 k_3) (k_2h_4 k_2) - 2 (k_3 h_1 k_3) (k_3 h_4 k_2)\\
L_2 &= 4 (k_2 h_1 h_4 k_3) - 2 (k_2 h_1 k_2) (k_2 h_4 k_3) - 2 (k_3 h_1 k_2) (k_3 h_4 k_3)\\
L_3 & = (k_3 h_1 k_3) (k_2 h_4 k_2)\\ L_4 & = (k_2 h_1 k_2) (k_3
h_4 k_3)\,\,.
\end{align*}
Let us now determine the residue of ${1}/{k_2k_3}$ that describes
the exchange of a massless particle, such as the graviton, the
dilaton or the antisymmetric rank 2 tensor. The Shapiro-Virasoro
factor gives 1 in the limit $k_2 k_3 \rightarrow 0$. Keeping
further in mind that the residue implies $k_2 k_3=0$, which
further implies via momentum conservation $k_2 =- k_3$ we obtain
\begin{align}
\label{eq residue}
\frac{2}{k_2 k_3}  v_1 \cdot v_4 \, (p_4 f_2 f_3 p_4) (h_1 h_2)  = \frac{2}{k_2 k_3}   (p_4 f_2 f_3 p_4) ({\cal H}_1 {\cal H}_4 )\,\,,
\end{align}
where we have reinstated the full projector ${\cal
H}_{i\mu_1\mu_2}= v_i \otimes h_{\mu_1\mu_2}$.

As mentioned above the residue describes the exchange of a
graviton, axion or dilaton. Note that the expression \eqref{eq
residue} does not contain any $\alpha'$ terms, they cancel among
each other once one imposes $k_2 =-k_3$. That is somewhat expected
since the on-shell coupling of the massless fields, graviton, axion
and dilaton, to the gauge bosons as well as to the massive BPS
states do not contain any $\alpha'$ corrections even for complex momenta.

In order to determine the couplings of HS states to a
specific massless field such as the graviton, we choose the gauge
bosons to have opposite helicity \cite{Feng:2010yx,
Bianchi:2008pu}. In this case the exchanged massless state is the
graviton since $T_{\mu\nu} = f^+_{\mu\rho}f^-_{\nu}{}^{\rho}$ is
symmetric and conserved. (Pseudo)scalars couple to
$f^+_{\mu\nu}f^{+\mu\nu} \pm f^-_{\mu\nu}f^{-\mu\nu}$. For
concreteness we choose positive helicity for $a_2= a^+_2 $ and
negative helicity for $a_3= a^-_3$. Then the residue \eqref{eq
residue} takes the form
\begin{align}
\frac{2}{k_2 k_3}  (p_4 f^+_2 f^-_3 p_1) ({\cal H}_1 {\cal H}_4 )
= \frac{2}{k_2 k_3}   \left(f^+_2 f^-_3\right)_{\mu\nu}
(p_1-p_4)^{\mu} ({\cal H}_1 {\cal H}_4 ) (p_1-p_4)^{\nu} \label{eq
residue before manipulation}
\end{align}
This structure is of the expected type
\begin{align}
{T^{(2,3)}}_{\mu\nu} {T^{(1,4)}}^{\mu\nu}
\end{align}
where
\begin{align}
{T^{(2,3)}}_{\mu\nu} = \left(f^+_2 f^-_3\right)_{\mu\nu} \qquad
\text{and} \qquad {T^{(1,4)}}^{\mu\nu} =  (p_1-p_4)^{\mu} ({\cal
H}_1 {\cal H}_4 ) (p_1-p_4)^{\nu}
\end{align}
denote the stress energy tensor of two gauge fields and two higher
spin fields, respectively.

In order to set the stage for the study of the low energy behavior
and spin effects, it proves convenient to express the
polarizations in terms of the momenta $p_i$ and $k_i$. Symmetric
tensor polarizations
\begin{align}
h^{\mu\nu}_1 &= \frac{1}{\sqrt{2}}  \left( w^{\mu}_2 w^{\nu}_3+ w^{\mu}_3 w^{\nu}_2\right) \hspace{32mm} \widetilde{h}^{\mu\nu}_1 = \frac{1}{\sqrt{2}}  \left( \widetilde{w}^{\mu}_2 \widetilde{w}^{\nu}_3+ \widetilde{w}^{\mu}_3 \widetilde{w}^{\nu}_2\right)\\
h^{\mu\nu}_2 &= \frac{1}{\sqrt{2}}  \left( w^{\mu}_3 w^{\nu}_1+ w^{\mu}_1 w^{\nu}_3\right)
\hspace{32mm} \widetilde{h}^{\mu\nu}_2 = \frac{1}{\sqrt{2}}  \left( \widetilde{w}^{\mu}_3 \widetilde{w}^{\nu}_1+ \widetilde{w}^{\mu}_1 \widetilde{w}^{\nu}_3\right)\\
h^{\mu\nu}_3 &= \frac{1}{\sqrt{2}}  \left( w^{\mu}_1 w^{\nu}_2+ w^{\mu}_2 w^{\nu}_1\right)
\hspace{32mm} \widetilde{h}^{\mu\nu}_3 = \frac{1}{\sqrt{2}}  \left( \widetilde{w}^{\mu}_1 \widetilde{w}^{\nu}_2+ \widetilde{w}^{\mu}_2 \widetilde{w}^{\nu}_1\right)\\
h^{\mu\nu}_4 &= \frac{1}{\sqrt{2}}  \left( w^{\mu}_1 w^{\nu}_1- w^{\mu}_3 w^{\nu}_3\right)
\hspace{32mm} \widetilde{h}^{\mu\nu}_4 = \frac{1}{\sqrt{2}}  \left( \widetilde{w}^{\mu}_1\widetilde{w}^{\nu}_1- \widetilde{w}^{\mu}_3 \widetilde{w}^{\nu}_3\right)\\
h^{\mu\nu}_5 &= \frac{1}{\sqrt{6}}  \left( w^{\mu}_1 w^{\nu}_1-2 w^{\mu}_2 w^{\nu}_2 + w^{\mu}_3 w^{\nu}_3\right)
\hspace{15mm} \widetilde{h}^{\mu\nu}_5 = \frac{1}{\sqrt{6}}  \left( \widetilde{w}^{\mu}_1 \widetilde{w}^{\nu}_1 -2 \widetilde{w}^{\mu}_2 \widetilde{w}^{\nu}_2 + \widetilde{w}^{\mu}_3 \widetilde{w}^{\nu}_3\right).
\end{align}
can be constructed from massive spin $1$ polarizations. The
massive spin $1$ polarizations $w_i$, transverse to $p_1$, are
given by
\begin{align}
w^{\mu}_1 &=\frac{1}{(k_2 k_3) \sqrt{-2F -M^2} } \epsilon^{\mu\nu\rho\sigma} {k_2}_{\nu}  {k_3}_{\rho}  {p_4}_{\sigma}\\
w^{\mu}_2 &=\frac{M} {(k_2 p_1) } k'^{\mu}_2\\
w^{\mu}_3 &=\frac{(k_2 p_1)}{(k_2 k_3) \sqrt{-2F -M^2} }  \left( k'^{\mu}_3 + \frac{(k'_3 k'_2)}{(k'_2 k'_2)} k'^{\mu}_3 \right)\,\,,
\end{align}
where $k'^{\mu}_i=\left( k^{\mu}_i + \frac{(k_i p_1)}{M^2}
p^{\mu}_1 \right)$ and
\begin{align}
F= \frac{(k_2 p_1)(k_3 p_1) }{k_2 k_3} \,\,.
\end{align}
The polarizations $\widetilde{w}_i$, transverse to $p_4$, are
given by $\widetilde{w}^{\mu}_1 = {w}^{\mu}_1$ and
\begin{align}
\widetilde{w}^{\mu}_2 &=\frac{M} {(k_2 p_4) } k''^{\mu}_2\\
\widetilde{w}^{\mu}_3 &=\frac{(k_2 p_4)}{(k_2 k_3) \sqrt{-2F -M^2}
} \left( k''^{\mu}_3 + \frac{(k''_3 k''_2)}{(k''_2 k''_2)}
k''^{\mu}_3 \right)
\end{align}
with $k''^{\mu}_i=\left( k^{\mu}_i + \frac{(k_i p_4)}{M^2}p^{\mu}_4 \right)$.

It is straightforward to show that $w_i$ and $\widetilde{w}_i$
represent two ortho-normal bases for space-like
vectors\footnote{Note that $-2F-M^2> 0$ for physical momenta.}.
This holds true also for the above bases of symmetric massive spin
$2$ polarizations $h^{\mu\nu}_i$ and $\widetilde{h}^{\mu\nu}_i$.

Given the above choice of tensor polarizations, one can proceed
analyzing specific amplitudes with fixed polarizations for the
higher spin states. It turns out that the polarizations $h_2$,
$h_3$ only couple to $\widetilde{h}_2$, $\widetilde{h}_3$. Here we
focus on this subset. Moreover we have to specify the helicity for
the two massless vector bosons, which can be written as
\cite{Choi:1994ax}
\begin{align}
a_2 = \frac{1}{\sqrt{2}} \left( n^{\mu}_1 + i \lambda_2 n^{\mu}_2
\right) \qquad \qquad   a_3 = \frac{1}{\sqrt{2}} \left( n^{\mu}_1
+ i \lambda_3  \,n^{\mu}_2 \right)\,\,,
\end{align}
where the $n_i$'s are given by
\begin{align}
n^{\mu}_1 &= \frac{1}{\sqrt{-2F -M^2}} \left[ (p^{\mu}_1-p^{\mu}_4) + \frac{(p_1 k_2) - (p_1 k_3)}{ (k_2 k_3)} (k^{\mu}_2 -k^{\mu}_3) \right] \\ n^{\mu}_2 &= \frac{1}{\sqrt{-2F -M^2}}  \frac{\epsilon^{\mu \nu \rho \sigma} {k_2}_{\nu} \, {k_3}_{\rho} \, {p_4}_{\sigma}   }{ k_2 k_3} \,\,.
\end{align}
Choosing opposite helicity for the vector boson polarizations one obtains
\begin{align}
\label{eq amplitude 22} {\cal A}^{h_2 \rightarrow \widetilde{h}_2}
&= 2\, {\cal A}^0 \,\,\left(1+ \frac{M^2}{2F} \right) \left( 1 -  \frac{ \alpha' (k_2 k_3)}{2} \left( 2+\frac{M^2}{F}\right) \right)\\
{\cal A}^{h_2 \rightarrow \widetilde{h}_3} &=2\, \cA^0 \,\, {\cal F}_{SV}\,\,
\left(1+ \frac{M^2}{2F} \right) \left(  \frac{ \alpha' (k_2
k_3)}{2} \frac{M}{F}\sqrt{-2F-M^2} \right)
\\
{\cal A}^{h_3 \rightarrow \widetilde{h}_2} &=2\,\cA^0 \,\,  \left(1+ \frac{M^2}{2F} \right) \left(  \frac{ \alpha' (k_2 k_3)}{2} \frac{M}{F}\sqrt{-2F-M^2} \right)\\
{\cal A}^{h_3 \rightarrow \widetilde{h}_3} &=2\, \cA^0 \,\, {\cal F}_{SV}\,\,
\left(1+ \frac{M^2}{2F} \right) \left( 1 +  \frac{ \alpha' (k_2
k_3)}{2} \frac{M^2}{F} \right) \,\,.\label{eq
amplitude 33}
\end{align}
with
\begin{align}
\cA^0= \,{g_{_{YM}}^2\over M_s^2}
(2\pi)^{4}\delta(\Sigma_ip_i) F  \,\, {\cal F}_{SV}\,\,.
\end{align}
Note that the diagonal terms, in which there is no polarization
flip, expose a ${1}/{k_2 k_3}$ pole indicating the exchange of a
graviton. On the other hand the off diagonal amplitudes do not
reveal such a pole. Thus there is no massless particle exchange
with  the latter choice of  polarizations for the massive spin 2
states. In addition the off-diagonal amplitudes contain square
roots of kinematical invariants, which seem unusual from a field
theoretic point of view. Generically one would expect tree-level
amplitudes to exhibit pure pole structure behavior. Note though
that these square roots scale with $\alpha'$ and disappear
in the field theory limit $\alpha' \rightarrow 0$. Thus they might
be thought to have stringy origin. Moreover these terms vanish in
the formal limit $M \rightarrow 0$. It should however be kept in mind
that `kinematical singularities' such as the above square-root
cuts are tolerable in field-theory amplitudes with massive
particles\footnote{M.~B. would like to thank G.~Veneziano for an
enlightening discussion on this point.}. In fact a preliminary
analysis of the other sector, corresponding to massive spin 2
polarizations $h_1, h_4, h_5$ mixing with $\tilde{h}_1,
\tilde{h}_4, \tilde{h}_5$, displays `kinematical singularities' to
leading order $\ap^0$ with suppression at least $(k_2k_3)/M$.

It should be clear that the limit $M \rightarrow 0$ can be only
taken in a formal sense, since after all the 1/2 BPS  spin $2$
states are genuinely massive $M > 2/\sqrt{\ap}$. The situation is
different for 1/2 BPS spin $1$ states which can become massless in
the interior of the moduli space. It is relatively easy to
investigate the amplitude involving two gauge bosons and two
massive 1/2 BPS spin $1$ states in connection to possible
`kinematical singularities'. Applying the general formula
\eqref{eq final result} to $s=1$ one obtains for the scattering of
two gauge bosons onto two massive spin $1$ states
\begin{align}
{\cal A} =-& {g_{_{YM}}^2\over M_s^2}
(2\pi)^{4}\delta(\Sigma_ip_i) \frac{1}{k_2k_3}  \,\, {\cal
F}_{SV}\,\, \delta^\perp_{ij} \, (p_4 f_2 f_3 p_4) \,\, \Big[
-(W^i_1\cdot W^j_4) + (k_2 W^i_1)(k_2 W^j_4) \\  & \hspace{2mm}   +(k_3
W^i_1)(k_3 W^j_4) - \frac{k_3 p_4 +1}{k_3 p_1} (k_3 W^i_1)(k_2
W^j_4) - \frac{k_3 p_1 +1}{k_3 p_4} (k_2 W^i_1)(k_3W^j_4) \Big]\,.
 \end{align}
Choosing opposite helicity for the gauge bosons we obtain the
following amplitudes
\begin{align}
 {\cal A}^{w_1 \rightarrow \widetilde{w}_1} &=- \cA^0\,\, {\cal F}_{SV}\,\, \left(1+ \frac{M_W^2}
{2F} \right) \\
 {\cal A}^{w_2 \rightarrow \widetilde{w}_2} &=\cA^0  \,\, {\cal F}_{SV}\,\,
\left(1+ \frac{M_W^2}{2F} \right) \left( 1+  \frac{ \alpha' (k_2 k_3)}{2}
\frac{M_W^2}{F}\right)\\
{\cal A}^{w_2 \rightarrow \widetilde{w}_3} &=-\cA^0\,\, {\cal F}_{SV}\,\,
\left(1+ \frac{M_W^2}{2F} \right)  \left(\frac{ \alpha' (k_2 k_3)}{2}
\frac{M_W}{F} \sqrt{-2F -M_W^2} \right)\\
{\cal A}^{w_3 \rightarrow \widetilde{w}_2} &=- \cA^0  \,\,
{\cal F}_{SV}\,\,  \left(1+ \frac{M_W^2}{2F} \right)
\left( \frac{ \alpha' (k_2 k_3)}{2} \frac{M_W}{F} \sqrt{-2F -M_W^2} \right)\\
{\cal A}^{w_3 \rightarrow \widetilde{w}_3} &=\cA^0\,\, {\cal F}_{SV}\,\,
\left(1+ \frac{M_W^2}{2F} \right) \left( 1 -  \frac{ \alpha' (k_2
k_3)}{2} \left( 2 +\frac{M_W^2}{F}\right) \right)\,\,.
 \end{align}
All other combinations give vanishing results. As for the massive
spin $2$ case, we encounter `kinematical singularities' in the
off-diagonal amplitudes that display a universal square-root
factor. These amplitudes scale to zero with $\alpha'$ and thus
disappear in the field theory limit $\alpha' \rightarrow 0$.
Moreover, in the limit of vanishing mass $M_W\rightarrow 0$ of the
1/2 BPS vector boson these terms vanish. That is exactly what one
expects since for scattering of four massless particles the
amplitude should not expose this kind of `kinematical
singularities'.

Let us return to the massive spin $2$ amplitude, and take the
formal limit $\alpha' \rightarrow 0$, keeping $M$ fixed. The
Shapiro-Virasoro factor ${\cal F}_{SV}$ behaves as
\begin{align}
{\cal F }_{SV} &= \frac{\Gamma\left[1 + \frac{\alpha'}{2} (k_2
k_3)\right]\,\, \Gamma\left[\frac{\alpha'}{2} ( k_3 p_4
)\right]\,\, \Gamma\left[ \frac{\alpha'}{2} (k_3 p_1)\right]}
{\Gamma\left[2 - \frac{\alpha'}{2} (k_2 k_3)\right]\,\,
\Gamma\left[-\frac{\alpha'}{2} ( k_3 p_4 )\right]\,\,
\Gamma\left[-\frac{\alpha'}{2} (k_3 p_1)\right]}\\
& \sim \frac{1}{1- \frac{\alpha'}{2} (k_2 k_3)}  \,\, \frac{\left[
1+ \frac{\alpha'}{2} (k_2 k_3 ) \psi(1) \right] \, \, \left[1 +
\frac{\alpha'}{2} (k_3 p_4) \psi(1) \right] \left[ 1 +
\frac{\alpha'}{2} (k_3 p_1) \psi(1) \right] }{\left[ 1-
\frac{\alpha'}{2} (k_2 k_3 ) \psi(1) \right] \, \, \left[1 -
\frac{\alpha'}{2} (k_3 p_4) \psi(1) \right] \left[ 1 -
\frac{\alpha'}{2} (k_3 p_1) \psi(1) \right]}
\\
& \sim [1 + \frac{\alpha'}{2} (k_2 k_3)] \left[ 1 + {\alpha'} (k_2
k_3 + k_2 p_1 + k_3 p_1) \psi(1) \right]  \sim 1 +
\frac{\alpha'}{2} (k_2 k_3) + ... \,\,.
\end{align}
due to momentum conservation.

In combination with the other parts of the amplitudes \eqref{eq
amplitude 22} to\eqref{eq amplitude 33} one obtains for the low
energy limit
\begin{align}
{\cal A}^{h_2 \rightarrow \widetilde{h}_2} &=2 \,{g_{_{YM}}^2\over
M_s^2} (2\pi)^{4}\delta(\Sigma_ip_i)  F \left[ \left(1+
\frac{M^2}{2F} \right) -  \frac{ \alpha' }{2} (k_2 k_3) \left(
1+\frac{M^2}{F}\right)^2 +{\cal O}\left(\frac{\alpha'}{2}\right)^2
 \right] \\
{\cal A}^{h_2 \rightarrow \widetilde{h}_3} &=2\, {g_{_{YM}}^2\over
M_s^2} (2\pi)^{4}\delta(\Sigma_ip_i)  \left[ \frac{ \alpha'}{2} \,
(k_2 k_3) \left(1+ \frac{M^2}{2F} \right) M \sqrt{-2F-M^2} +{\cal
O}\left(\frac{\alpha'}{2}\right)^2\right]
\\
{\cal A}^{h_3 \rightarrow \widetilde{h}_2} &=2\, {g_{_{YM}}^2\over
M_s^2} (2\pi)^{4}\delta(\Sigma_ip_i)  \left[ \frac{ \alpha'}{2} \,
(k_2 k_3) \left(1+ \frac{M^2}{2F} \right) M \sqrt{-2F-M^2} +{\cal
O}\left(\frac{\alpha'}{2}\right)^2\right]
\\
{\cal A}^{h_3 \rightarrow \widetilde{h}_3} &=2\, {g_{_{YM}}^2\over
M_s^2} (2\pi)^{4}\delta(\Sigma_ip_i) F \left[ \left(1+ \frac{
M^2}{2F} \right) + \frac{ \alpha' }{2}  (k_2 k_3)    \left(
1+\frac{M^2}{F}\right)^2 +{\cal O}\left(\frac{\alpha'}{2}\right)^2
 \right]
\end{align}

One can also consider the Regge limit of the above amplitudes that
corresponds to taking $\ap(s-M^2) >> 1$ with $t$ fixed. In this
limit the Shapiro Virasoro factor scales like \be \cF_{SV} \approx
{\Gamma\left(1-{\ap \over 4}t\right) \over \Gamma\left(2+{\ap
\over 4}t\right)} e^{+i\pi \ap t/4} \left( \ap {s- M^2 \over
4}\right)^{\ap t/2} + ...\ee One can recognize in order the SV
`form factor', a phase shift and a power law suppression ($t<0$ in
the physical domain). The latter suggest the possibility that
higher loop contributions exponentiate in the eikonal
approximation.  Contrary to what happens for the scattering of
gravitons off D-branes \cite{D'Appollonio:2010ae}, the lack of a
valid semi-classical limit for 1/2 BPS HS states seems not to
favor this interpretation here. Clearly this point deserves
further study.

Another interesting limit is the limit in which  momentum exchange
in the $t$-channel is extremely small. To analyze this limit it is
convenient to work in the Lab frame, where\footnote{We take all
the momenta to be incoming, thus one has to remember that
$k_{3,4}^{phys}=-k_{3,4}$. Henceforth we assume $E_{3,4}$ to be
the physical energies.} $p_1=(M,{\bf 0})$, $k_2=(E_2,{\bf k}_2)$,
$-k_3=(E_3,{\bf k}_3)$ and $-p_4=(E_4,{\bf p}_4)$. With that
choice the Mandelstam variables assume values \beqn
&&s=-(k_2+p_1)^2=M(M+2E_2)\\
\label{tLAB}
&&t=-(k_2+k_3)^2=-2E_2E_3(1-\cos\theta)\\
&&u=-(k_3+p_1)^2=M(M-2E_3) \,\,.\eeqn The angular dependance in  $t$
can be eliminated by means of the relation $s+t+u=2M^2$ leading to
\beqn \label{cosangle} 1-\cos\theta=\frac{M(E_2-E_3)}{E_2E_3}
\eeqn so that (\ref{tLAB}) becomes $t=-2M(E_2-E_3)$. Let us
introduce the variables $Q^2=-t\geq 0$ and $M E= 2 M (E_2+E_3)$.
In terms of these new variables and reinstating $\alpha'$ factors
the Shapiro-Virasoro factor \eqref{eq S-V factor} reads \beqn
\label{S-VFormFact} \cF_{SV}=\frac{\Gamma(1+\frac{\alpha'}{2}
\frac{Q^2}{2})\Gamma(-\frac{\alpha'}{2}\frac{Q^2+EM}{4})\Gamma(-\frac{\alpha'}{2}
\frac{Q^2-EM}{4})}{\Gamma(2-\frac{\alpha'}{2}\frac{Q^2}{2})\Gamma(\frac{\alpha'}{2}
\frac{Q^2+EM}{4})\Gamma(\frac{\alpha'}{2} \frac{Q^2-EM}{4})} \eeqn In the
limit $Q^2 \rightarrow 0$ the Shapiro Virasoro factor scales like
\begin{align}
\lim_{Q^2 \rightarrow 0} {\cal F}_{SV} \sim 1 + \frac{\alpha'}{2} Q^2 \left[\frac{1}{2}-\gamma-\psi\left(\frac{\alpha' EM}{8}\right)-\frac{4}{\alpha' EM}-\frac{\pi}{2}\cot\left(\frac{\pi \alpha' EM}{8}\right)  + {\cal O} (Q^2) \right]\,\,,
\end{align}
where $\psi (z) = \partial \log \Gamma (z)$ and $\gamma$ is the Euler-Mascheroni constant.
Combining this with the remaining part of the diagonal amplitudes gives
\begin{align}
{\cal A}^{h_2 \rightarrow \widetilde{h}_2} &=\frac{1}{4} \,{g_{_{YM}}^2\over
M_s^2} (2\pi)^{4}\delta(\Sigma_ip_i)  \frac{M^2 E^2 }{Q^2}
\left[ 1 + Q^2 \left(-\frac{4}{E^2} \right) \right. \\
& \left.\hspace{10mm}+ \frac{ \alpha'  }{2}  Q^2\left(-\frac{1}{2} -\gamma  -\psi\left(\frac{\alpha' EM}{8}\right)-\frac{\pi}{2}\cot\left(\frac{\pi \alpha' EM}{8}\right) \right)  + {\cal O} (Q^2)
 \right] \\
 {\cal A}^{h_3 \rightarrow \widetilde{h}_3} &=\frac{1}{4}\, {g_{_{YM}}^2\over
M_s^2} (2\pi)^{4}\delta(\Sigma_ip_i) \frac{M^2 E^2}{Q^2}
\left[ 1 + Q^2 \left(  -\frac{4}{E^2} \right) \right.
\\
&  \left.\hspace{10mm}
 + \frac{ \alpha'  }{2} Q^2\left(\frac{1}{2} -\gamma  -\psi\left(\frac{\alpha' EM}{8}\right)-\frac{\pi}{2}\cot\left(\frac{\pi \alpha' EM}{8}\right) \right)
 + {\cal O} (Q^2)  \right] \,\,.
\end{align}
As expected we find $\alpha' Q^2$ corrections to the massless
exchange, due to the infinite tower of string excitations. However
even in the low energy limit $\alpha' \rightarrow 0$,  there is a
$Q^2$ correction to the ``minimal" coupling of the higher spin
states to the graviton that results from a conspiracy between
factors of $\ap$ in the numerator and in the denominator and may
be interpreted as a remnant of the non-locality of string
interactions.

\section{Conclusions}
\label{Final}

Let us conclude by summarizing our results and mentioning open
problems. By studying physical processes involving
(non)-perturbatively stable 1/2 BPS higher spin states in $\cN=4$
toroidal compactifications of the heterotic string we have gained
some insights into their dynamical properties.

Similar results are expected to hold for $\cN=2$
compactifications. In particular our analysis carries over
immediately to freely acting orbifolds such as the FHSV model
\cite{Ferrara:1995yx, Klemm:2005pd} that are expected to receive
no quantum corrections to the two-derivative effective action,
since $N_h = N_v$ everywhere in the moduli space. Meta-stability
of higher spin states is expected in $\cN=1$ compactifications
with `Large Extra Dimensions' \cite{ArkaniHamed:1998rs,
Antoniadis:1998ig}, that are difficult to accommodate in the
heterotic string though \cite{Benakli:1999yc}.

We have also derived explicit expressions for massive non BPS HS
states in the first and second massive levels, that couple to
pairs of BPS states. Finally, we have analyzed some scattering
processes involving BPS HS states, paying attention to spin
effects and low-energy limits. Contrary to the recently analyzed
case of scattering of gravitons on D-branes
\cite{D'Appollonio:2010ae} for which the eikonal approximation
allows to reconstruct the `classical' geometry, in our case
amplitudes are resilient to such a semi-classical approximation,
since no supergravity solution can account for 1/2 BPS states with
(higher) spin \cite{Elvang:2004rt, Bena:2004de, Dabholkar:2006za,
Bena:2007kg, Iizuka:2007sk}. This seems to raise a puzzle in the
identification of the microstates accounting for the entropy of
small BH's with two charges \cite{Sen:2009bm, Dabholkar:2010rm}.
We have pointed out that the only protected quantity, the helicity
super-trace $\cB_4 = Str (2h)^4$, makes no distinction between
$(2s+1)$ vector multiplets and one spin $s$ supermultiplet.
Further analyses of the thermo-dynamical properties of these
peculiar HS states is definitely necessary in order to clarify the
situation.

The tree-level scattering amplitude of gauge bosons on higher spin
states, we studied, exhibits a massless pole and with the
appropriate helicity choice for the gauge bosons we extracted the
coupling of HS states to the graviton. Equipped with a convenient
basis for the polarizations of the HS states, we have investigated
interesting limits of the amplitude. For particular choices of the
projections the amplitudes reveal unusual `kinematical
singularities', which deserve further study. Moreover, it would be
interesting to extend this study to higher orders in perturbation
theory and to other HS states, be they BPS and thus stable or not.

 \section*{Acknowledgments}

Discussions with S.~Aoyama, M.~Bochicchio, G.~D'Appollonio,
S.~Ferrara, D.~Forcella, C.~Kounnas, J.~F.~Morales, F.~Riccioni,
A.~Sagnotti, O.~Schlotterer, A.~Sen, M.~Taronna, T.~Taylor,
G.~Veneziano are kindly acknowledged. This work was partially
supported by the ERC Advanced Grant n.226455 {\it ``Superfields''}
and by the Italian MIUR-PRIN contract 2007-5ATT78 {\it
``Symmetries of the Universe and of the Fundamental
Interactions''}.


\appendix

\section{Higher Spin 1/2 $BPS$ $\cN=2$ super-multiplets
in $D=4$
\label{app higher spins in N=2}}

As in the $\cN=4$ supersymmetric case discussed in Section
\ref{BPSHS}, in $\cN=2$ supersymmetric compactifications of the
Heterotic String, such as the FHSV model \cite{Ferrara:1995yx,
Klemm:2005pd}, the number of complex charged states in $1/2$ BPS
multiplets with $s\neq 0$ can be easily obtained from the vertex
operators \beqn V^{(-1)}_{\cU_I} &=& \cU_{I,\mu_1 \cdots \mu_s}
e^{-\varphi} \psi^I e^{i{\bf p}_L {\bf X}_L} \bar\de X_R^{\mu_1}
\cdots\bar\de X_R^{\mu_s} e^{i{\bf p}_R {\bf X}_R} e^{i p X} \eeqn
with internal excitations $I=1,2$ (untwisted directions only!) in
the ground state of the L-moving sector, and \beqn
V^{(-1)}_{\cU_\mu} &=& \cU_{\mu,\mu_1 \cdots \mu_s} e^{-\varphi}
\psi^\mu e^{i{\bf p}_L {\bf X}_L} \bar\de X_R^{\mu_1}
\cdots\bar\de X_R^{\mu_s} e^{i{\bf p}_R {\bf X}_R} e^{i p X} \eeqn
with space-time excitations in the ground state of the L-moving sector.\\
The tensors $\cU_{I,\mu_1... \mu_s}$ and $\cU_{\mu,\mu_1...
\mu_s}$ are totally symmetric by construction in the $\mu_i$
indices and, in order for the states to be BRST invariant, they
should satisfy \beqn \label{BRSBHcond2} p_L^I\cU_{I,\mu_1 \cdots
\mu_s}=p^\mu\cU_{I,\mu\mu_2 \cdots\mu_s}=\eta^{\mu\nu} \cU_{I,\mu\nu\mu_3 \cdots\mu_s}=0\\
p^\mu\cU_{\mu,\mu_1 \cdots \mu_s}=p^{\mu_1}\cU_{\mu,\mu_1\mu_2
\cdots\mu_s}=\eta^{\mu_1\mu_2}\cU_{\mu,\mu_1\mu_2\mu_3
\cdots\mu_s}=0 \eeqn The tensor $\cU_{I,\mu_1... \mu_s}$ accounts
for one\footnote{After imposing BRST conditions (\ref{BRSBHcond2})
the internal index $I$ is allowed to run only over the single
`untwisted' direction orthogonal to the central charge $p^I_L$.}
states of spin $s$, while $\cU_{\mu,\mu_1... \mu_s}$ gives rise to
spin $s+1$, $s$ and $s-1$ states. Therefore the number of bosonic
degrees of freedom is \beqn
n_B=2(s+1)+1+(1+1)(2s+1)+2(s-1)+1=(2s+1)4_B \eeqn

Vertex operators for the fermionic states read \beqn
V^{(-1/2)}_{\Upsilon_{\a r}} &=& \Upsilon_{\a r,\mu_1 \cdots
\mu_s} e^{-\varphi/2} S^\a S^r \Sigma_+ e^{i{\bf p}_L {\bf X}_L}
\bar\de X_R^{\mu_1} \cdots\bar\de X_R^{\mu_s} e^{i{\bf p}_R {\bf
X}_R} e^{i p X} \eeqn and \beqn V^{(-1/2)}_{\ov\Upsilon_{\dot\a
\dot{r}}} &=& \ov\Upsilon_{\dot\a \dot{r},\mu_1 \cdots \mu_s}
e^{-\varphi/2} C^{\dot\a} C^{\dot{r}} \Sigma_-
 e^{i{\bf p}_L {\bf X}_L} \ov\de X_R^{\mu_1}
\cdots\ov\de X_R^{\mu_s} e^{i{\bf p}_R {\bf X}_R} e^{i p X} \eeqn
where $S^\a$, $C^{\dot\a}$ are $SO(3,1)$ spin fields and
$C^{{r}}$, $ C^{\dot{r}}$ are $SU(2)$ spin fields. BRST invariance
requires \be p^\mu \ov\sigma_\mu^{\dot\a \a} \Upsilon_{\a r,\mu_1
\cdots \mu_s} + p_{I,L}\tau^I_{rs}\ov\Upsilon^{s\dot\a}_{\mu_1
\cdots \mu_s} = 0 \ee that allows to express
$\ov\Upsilon^{\dot\a\dot{r}}_{\mu_1 \cdots \mu_s}$ in terms of
$\Upsilon_{\a r,\mu_1 \cdots \mu_s}$ \be
\ov\Upsilon^{A\dot\a}_{\mu_1 \cdots \mu_s} = {1\over M^2} p^\mu
\ov\sigma_\mu^{\dot\a \a} p^I_{L}\ov\tau_I^{rs}\ov\Upsilon_{\a
s,\mu_1 \cdots \mu_s} \ee with $M^2= -p\cdot p = |\bf p_L|^2$. Combing
spin 1/2 from left-movers with spin $s$ from Right-movers one gets
$s+1/2$ and $s-1/2$. Taking into account the degeneracy ${\bf 2}$
of $SU(2)$, the number of fermionic degrees of freedom turns out
to be \beqn n_F=2[2(s+1/2)+1]+2[2(s-1/2)+1]=(2s+1)4_F \eeqn Thus,
these multiplets contains $(2s+1)(4_B-4_F)$ complex charged states
with maximal spin $s_{hws}=s+1$ and minimal spin $s_{lws}=s-1$.

Similar 1/2 BPS HS $\cN = 2$ multiplets arise by combining
Left-movers `odd' under twist (`odd' ground states) with
Right-movers `odd' under twist. In $\cN = 1$ no particle-like 1/2
BPS multiplet is possible. Non BPS $\cN = 1$ HS multiplets consist
in one spin $s$, two spin $s-1/2$ and one spin $s-1$ \cite{Zinoviev:2007ig}.

\section{Details of the amplitude calculation
\label{app details of amplitude}}

Here we display all the necessary correlators to compute the
4-point amplitudes in Section \ref{4-ptamp}. Let us
start with the left-moving part of the amplitude \eqref{eq 4 point amplitude internal} given by equation \eqref{eq left-moving part
4 point st}.
Using standard OPE's one gets for the left-moving correlators
\begin{align}
& \langle e^{i {\bf p}^1_L { X}}(z_1) e^{i {\bf p}^4_L {X}}(z_4)\rangle =
z^{{\bf p}^1_L\,{\bf p}^4_L}_{14}= z^{-|{\bf p}^1_L|^2}_{14} = z^{-M^2}_{14}
\\ & \hspace{0mm}  \langle \psi^{i}(z_1) \, \psi^{j}(z_4) \rangle =
\frac{\delta^{ij}}{z_{14}} \qquad \qquad
 \langle e^{-\varphi(z_1)} e^{-\varphi(z_4)}\rangle =z^{-1}_{14}
\end{align}
and
\begin{align}
&\Big\langle e^{ip_1 X}(z_1) \left[\partial X^{\kappa} - i (k_2 \cdot \psi)
\psi^{\kappa}\right]e^{ik_2 X}(z_2)\left[\partial X^{\lambda} - i (k_3 \cdot \psi) \right]\psi^{\lambda}
)e^{ik_3X}(z_3) \,e^{ip_4{X}}(z_4)\Big\rangle\\ & =\frac{\prod_{i\neq j} z^{k_i k_j}_{ij}}{z^2_{23} z_{14}} \,\,\left[ \eta^{\kappa \lambda} (k_2 \cdot k_3 -1) -  \frac{z_{14}\,z_{23}}{z_{12}\, z_{34}}    k^{\kappa}_3 p^{\lambda}_4+ \frac{z_{14}\,z_{23}}{z_{13}\, z_{24}}  p^{\kappa}_{4} k^{\lambda}_{2} -\frac{z^2_{14}\, z^2_{23}}{z_{12}\, z_{13}\, z_{24}\, z_{34} }  p^{\kappa}_{4} p^{\lambda}_{4}   \right]\,\,.
\label{eq F_i}
\end{align}

The right-moving part given in equation \eqref{eq right-moving
part 4-point st} contains the correlators
\begin{align}
\Big\langle \ov J^{a_2}(\ov z_2) \ov J^{a_3}(\ov z_3) \Big\rangle
= \frac{\delta^{a_2a_3}}{\ov z^2_{23}} \qquad \Big \langle
e^{i{\bf p}^1_R { X}(\ov{z}_1)} e^{i{\bf p}^4_R X(\bar{z}_4)
}\Big\rangle =\bar z^{{\bf p}^1_R {\bf p}^4_R}_{14} = \bar
z^{-|{\bf p}_R|^2}_{14} = \bar z^{2s-2-M^2}_{14}
\end{align}
and the correlator
\begin{align}
&\Big\langle\bar\de
X^{\mu_1} \cdots\bar\de X^{\mu_s}  e^{i p_1
X} (\bar z_1) \, e^{i k_2
X} (\bar z_2) \, e^{i k_3
X} (\bar z_3) \,
\bar\de
X^{\nu_1} \cdots\bar\de X^{\nu_s}  e^{i p_4
X} (\bar z_4)\Big\rangle \\
& \hspace{7mm} =\frac{\prod_{i\neq j} \bar z^{k_i k_j}_{ij}}{\bar z^{2s}_{14}} \sum^s_{k=0}
\left(\begin{array}{c}
s  \\
k
\end{array}\right)^2 (s-k)!\, (-1)^{s-k} \,\eta^{\mu_{k+1}\nu_{k+1}} ...  \eta^{\mu_{s}\nu_{s}}  \\ \nonumber
& \hspace{7mm}\qquad \qquad \times
\sum^k_{n=0} \, \sum^{k-2n}_{m=-n} \frac{k!\,\, \left(K^{\mu\nu}_1\right)^{k-2n-m} \, \left(K^{\mu\nu}_2\right)^{n} \, \left(K^{\mu\nu}_3\right)^{m+n}}{(k-2n-m)!\,\, n! \,\, (n+m)!}   \left( \frac{\bar z_{12}
\bar z_{34}}{\bar z_{13} \bar z_{24}}\right)^m
\end{align}
whose derivation is presented in appendix \ref{app correlator}. Here the $K_i$'s are given by
\begin{align}
K^{\mu\nu}_1= k^{\mu}_2\, k^{\nu}_2 + k^{\mu}_3\, k^{\nu}_3 \qquad K^{\mu\nu}_2= k^{\mu}_2\, k^{\nu}_3\qquad K^{\mu\nu}_3= k^{\mu}_3\, k^{\nu}_2\,\,.
\end{align}

Combining the left-moving and right-moving part and fixing three vertex operator positions to
\begin{align}
z_1 =z_{\infty}=\infty \qquad z_2 =1 \qquad z_3 =z  \qquad z_4 =0
\end{align}
and  after including the $c$-ghost contribution
\begin{align}
\Big\langle c(z_1)\, c(z_2) c(z_4)\Big\rangle \,\, \Big\langle  \ov c(\ov z_1)\, \ov c(\ov z_2) \ov c(\ov z_4)\Big\rangle = |z_{12}|^2
|z_{14}|^2  |z_{24}|^2
\end{align}
 we obtain
\begin{align}
&{\cal A}= {v_1}_i \otimes {h_1}_{\mu_1...\mu_s } {a_2}_{\kappa}   {a_3}_{\lambda} {v_4}_{j} \otimes {h_4}_{\nu_1...\nu_s }  \delta^{ij}
\int d^2 z \, |z|^{2k_3 p_4} |1-z|^{2 k_2 k_3 -4} \\
& \hspace{6mm}\times \left[ \eta^{\kappa \lambda} (k_2 \cdot k_3 -1) -  \frac{z_{14}\,z_{23}}{z_{12}\, z_{34}}    k^{\kappa}_3 p^{\lambda}_4+ \frac{z_{14}\,z_{23}}{z_{13}\, z_{24}}  p^{\kappa}_{4} k^{\lambda}_{2} -\frac{z^2_{14}\, z^2_{23}}{z_{12}\, z_{13}\, z_{24}\, z_{34} }  p^{\kappa}_{4} p^{\lambda}_{4}   \right]  {\cal J}(\ov z)
\end{align}
with
\begin{align}
{\cal J}(\ov z) &=\sum^s_{k=0}
\left(\begin{array}{c}
s  \\
k
\end{array}\right)^2 (s-k)!\, (-1)^{s-k} \,\eta^{\mu_{k+1}\nu_{k+1}} ...  \eta^{\mu_{s}\nu_{s}}  \\ \nonumber
& \hspace{7mm}\qquad \qquad \times
\sum^k_{n=0} \, \sum^{k-2n}_{m=-n} \frac{k!\left(K^{\mu\nu}_1\right)^{k-2n-m}  \left(K^{\mu\nu}_2\right)^{n}  \left(K^{\mu\nu}_3\right)^{m+n}}{(k-2n-m)!\,\, n! \,\, (n+m)!}   \ov z^m
\end{align}
After a few manipulations the amplitude can be written in manifestly gauge invariant form
\begin{align}
{\cal A}=  - {h_1}_{\mu_1...\mu_s }\, {h_4}_{\nu_1...\nu_s }
\int d^2 z \, |z|^{2k_3 p_4} |1-z|^{2 k_2 k_3 -4}  \frac{k_2k_3-1}{(k_3p_1)(k_3 p_4)}  v_1 \cdot v_4 \, (p_4 f_2 f_3 p_4) {\cal J}(\ov z)\,\,,
\end{align}
where $f_i$ denotes the field strength
\begin{align}
{f_i}^{\mu\nu} = k^{\mu}_i a^{\nu}_i -  a^{\mu}_i k^{\nu}_i\,\,.
\end{align}
Finally, with the integral
\begin{align}
\int d^2 z |z|^{2\,k_3\,p_4} |1-z|^{2\, k_2 \,k_3-4} \, \bar z^m &= \frac{(k_3 \, p_1)(k_3 \, p_4)}{(k_2\,k_3-1)(k_2\, k_3)}\,\, \frac{\Gamma(1+k_2\,k_3)\Gamma(k_3\, p_4)\, \Gamma(k_3\, p_1)}{\Gamma(2-k_2\, k_3)\Gamma(-k_3\, p_4)\, \Gamma(-k_3\, p_1)} \,\, J_m
\end{align}
where $J_m$ is given by\footnote{Note that $J_0=1$ }
\beqn
J_{m}=(-1)^m \prod_{r=1}^{|m|}\frac{\theta(m)\,(k_3\,p_4)+\theta(-m)\, (k_3\,p_1)+r}{\theta(m)\, (k_3\, p_1)+\theta(-m)\, (k_3\,p_4)+1-r}
\eeqn
one obtains
\begin{align}
{\cal A} =- \frac{1}{k_2k_3}   v_1 \cdot v_4 \, (p_4 f_2 f_3 p_4)     {h_1}_{\mu_1...\mu_s }\, {h_4}_{\nu_1...\nu_s } \,  {\cal F}_{SV}\,  {\cal J}^{\mu_1...\mu_s\nu_1...\nu_s}_R
\end{align}
with
\begin{align} \label{eq final equation for J}
{\cal J}^{\mu_1...\mu_s\nu_1...\nu_s}_R &=  \sum^s_{k=0}
\left(\begin{array}{c}
s  \\
k
\end{array}\right)^2 (s-k)!\, (-1)^{s-k} \,\eta^{\mu_{k+1}\nu_{k+1}} ...  \eta^{\mu_{s}\nu_{s}}  \\ \nonumber
& \hspace{7mm}\qquad \qquad \times
\sum^k_{n=0} \, \sum^{k-2n}_{m=-n} \frac{k!\left(K^{\mu\nu}_1\right)^{k-2n-m}  \left(K^{\mu\nu}_2\right)^{n}  \left(K^{\mu\nu}_3\right)^{m+n}}{(k-2n-m)!\,\, n! \,\, (n+m)!}   J_m
 \end{align}
 and the Shapiro-Virasoro factor
 \begin{align}
 {\cal F}_{SV} =
\frac{\Gamma(1+k_2\,k_3)\Gamma(k_3\,p_4)\Gamma(k_3\,p_1)}
{\Gamma(2-k_2\,k_3)\Gamma(-k_3\,p_4)\Gamma(-k_3\,p_1)}
\,\,.
 \end{align}

\section{ HS correlator
\label{app correlator}}

Here we present the derivation of the correlator
\begin{align}
\label{eq correlator higher spin}
{h_1}_{\mu_1 ... \mu_s} \,{h_4}_{\nu_1 ... \nu_s} \,\,
\Big\langle\bar\de
X^{\mu_1} \cdots\bar\de X^{\mu_s}  e^{i p_1
X} \, e^{i k_2
X} \, e^{i k_3
X} \,
\bar\de
X^{\nu_1} \cdots\bar\de X^{\nu_s}  e^{i p_4
X} \Big\rangle
\end{align}
for an arbitrary $s$ assuming that the polarization rank $s$ tensors $h_1$ and $h_4$ are completely symmetric. Rewriting the products of $\ov \partial X^{\mu_k}$ by
\beqn
\ov \de
X^{\mu_1} ... \,\ov \de X^{\mu_s}=\left[\frac{\ov \de}{\ov \de\alpha_{\mu_1}}... \, \frac{\ov \de}{\ov \de\alpha_{\mu_s}} \ e^{\alpha\ov\de X} (\ov z)\right]_{\vec{\alpha}=0}
\eeqn
and using the usual OPE's  the correlator \eqref{eq correlator higher spin} becomes
\beqn
\label{eq HR1}
H_R=\cI_R(p_i,\bar{z}_i)\left[\frac{\de}{\de\alpha_{\mu_1}}\cdots\frac{\de}{\de\alpha_{\mu_s}}\frac{\de}{\de\beta_{\nu_1}}\cdots\frac{\de}{\de\beta_{\nu_s}} \ exp\left(-\frac{\alpha\beta}{\bar{z}_{14}^2}-i\sum_{n\neq 1}\frac{\alpha p_n}{\bar{z}_{1n}}-i\sum_{m\neq 4}\frac{\beta p_m}{\bar{z}_{4m}}\right)\right]_{^{\alpha=0}_{\beta=0}}
\eeqn
where  $\cI_R(p_i,
\bar{z}_i) = \prod_{i<j} \bar{z}_{ij}^{ {{}}p_i p_j}$ is the
Koba-Nielsen factor for the right-moving part. Introducing the variables
\beqn
\widetilde{\alpha}^\mu=\alpha^\mu+\alpha^\mu_0, \qquad \qquad \qquad  \widetilde{\beta}^\nu=\beta^\nu+\beta^\nu_0
\eeqn
with
\begin{align}
\alpha^\mu_0=i\bar{z}_{14}^2\sum_{m\neq 4}\frac{p^\mu_m}{\bar{z}_{4m}} \qquad \qquad \qquad \beta^\nu_0=i\bar{z}_{14}^2\sum_{n\neq 1}\frac{p^\nu_n}{\bar{z}_{1n}}
\end{align}
allows for the compact form
\beqn
\cI_R(p_i,\bar{z}_i) \ exp\left(-\bar{z}_{14}^2\sum_{^{n\neq 1}_{m\neq 4}}\frac{p_np_m}{\bar{z}_{1n}\bar{z}_{4m}}\right)\left[\frac{\de}{\de\tilde{\alpha}_{\mu_1}}\cdots\frac{\de}{\de\tilde{\alpha}_{\mu_s}}\frac{\de}{\de\tilde{\beta}_{\nu_1}}\cdots\frac{\de}{\de\tilde{\beta}_{\nu_s}} \ e^{-\tilde{\alpha}\tilde{\beta}/\bar{z}_{14}^2}\right]_{^{\tilde{\alpha}=\alpha_0}_{\tilde{\beta}=\beta_0}}
\eeqn
Performing the derivatives one gets
\begin{align}
\label{HR2}
&\cI_R(p_i,\bar{z}_i)\bar{z}_{14}^{-2s}\sum_{k=0}^s\binom{s}{k}^2  (s-k)!\, (-1)^{s-k} \,\eta^{\mu_{k+1}\nu_{k+1}} ...  \eta^{\mu_{s}\nu_{s}}   \\ \nn
& \hspace{20mm} \times \left(\sum_{n\neq 1}\frac{p^{\mu_{1}}_n}{\ov{z}_{1n}}  \ov z_{14}\right)\cdots\left(\sum_{r\neq 1}\frac{p^{\mu_k}_r}{\ov{z}_{1r}}  \ov z_{14}\right)
 \left(\sum_{m\neq 4}\frac{p^{\nu_{1}}_m}{\ov{z}_{4m}}  \ov z_{14}\right)\cdots\left(\sum_{q \neq 4}\frac{p^{\nu_{k}}_q}{\ov{z}_{4s}} \ov z_{14}\right)
\end{align}
Here we already used the fact that the correlator is symmetric
under the exchange of $\mu_i \leftrightarrow \mu_j$ and  $\nu_i
\leftrightarrow \nu_j$. Now using the BRST constraint $p^{\mu_j}_i
h_{\mu_1 ... \mu_j ... \mu_s}=0$ and momentum conservation we can
manipulate this to
\begin{align}
\cW^{s-t}_R (\bar z_i) &= \frac{\prod_{i\neq j} \bar z^{k_i k_j}_{ij}}{\ov z^{2s}_{14}} \sum^s_{k=0}
\binom{s}{k}^2 (s-k)!\, (-1)^{s-k} \,\eta^{\mu_{k+1}\nu_{k+1}} ...  \eta^{\mu_{s}\nu_{s}}  \\ \nonumber
& \qquad \qquad \times
\sum^k_{n=0} \, \sum^{k-2n}_{m=-n} \frac{k!\,\, \left(K^{\mu\nu}_1\right)^{k-2n-m} \, \left(K^{\mu\nu}_2\right)^{n} \, \left(K^{\mu\nu}_3\right)^{m+n}}{(k-2n-m)!\,\, n! \,\, (n+m)!}   \left( \frac{\bar z_{12}
\bar z_{34}}{\bar z_{13} \ov z_{24}}\right)^m
\end{align}
where
\begin{align}
K^{\mu\nu}_1= k^{\mu}_2\, k^{\nu}_2 + k^{\mu}_3\, k^{\nu}_3 \qquad K^{\mu\nu}_2= k^{\mu}_2\, k^{\nu}_3\qquad K^{\mu\nu}_3= k^{\mu}_3\, k^{\nu}_2\,\,.
\end{align}
 Again we take advantage of the fact that the correlator is symmetric under the exchange of  $\mu_i \leftrightarrow \mu_j$ and  $\nu_i \leftrightarrow \nu_j$. The product of the respective powers of $K^n_i$ are understood as displayed below
\begin{align}
\left(K^{\mu\nu}_1\right)^{\alpha} \, \left(K^{\mu\nu}_2\right)^{\beta} \, \left(K^{\mu\nu}_3\right)^{\gamma}  =
\prod^{\alpha}_{u=1}   K^{\mu_u\nu_u}_1 \prod^{\alpha +\beta}_{v=\alpha+1} K^{\mu_v \nu_v}_2 \prod^{\alpha +\beta+\gamma}_{x=\alpha+\beta+1} K^{\mu_x \nu_x}_3 \,\,.
 \end{align}

\newpage


\providecommand{\href}[2]{#2}\begingroup\raggedright\endgroup
\end{document}